\documentclass[11pt]{article}
\usepackage[letterpaper,margin=1in]{geometry}
\usepackage{setspace}
\usepackage[T1]{fontenc}
\usepackage{baskervillef}
\usepackage[varqu,varl,var0]{inconsolata}
\usepackage[scale=.95,type1]{cabin}
\usepackage[baskerville,vvarbb]{newtxmath}
\usepackage[cal=boondoxo]{mathalfa}
\usepackage{latexsym}

\usepackage{amssymb}

\usepackage{amsthm}
\usepackage{amscd}
\usepackage{amsmath}
\usepackage{thm-restate}
\usepackage{tikz-cd}
\usepackage{tikz-3dplot}
\usepackage[all]{xy}
\usepackage{tabularx}
\usepackage{caption}
\usepackage{subcaption}
\usepackage{pgfplots}
\usepackage{algorithm}
\usepackage{algpseudocode}
\usepackage{algorithmicx}
\usepackage{hyperref}
\usepackage{cleveref}
\usepackage{float}
\usepackage{diagbox}
\usepackage{bm}
\usepackage{enumitem}
\usepackage{mathtools}
\floatname{algorithm}{Problem}
\algrenewcommand\algorithmicrequire{\textbf{Input:}}
\algrenewcommand\algorithmicensure{\textbf{Output:}}
\let\oldReturn\Return
\renewcommand{\Return}{\State\oldReturn}

\usetikzlibrary{decorations.markings}

\hypersetup{
  colorlinks   = true, 
  urlcolor     = blue, 
  linkcolor    = blue, 
  citecolor   = red 
}
\input xy \xyoption{frame}
\xyoption{dvips}

\usetikzlibrary{chains,scopes,decorations.markings}
\usepackage{subfiles}

\theoremstyle{definition}
\newtheorem{claim}{Claim}
\newtheorem{theorem}{Theorem}
\newtheorem{definition}{Definition}

\newtheorem{proposition}[theorem]{Proposition}
\newtheorem{observation}{Observation}

\newenvironment{cproof}
{\begin{proof}
 [Proof of Claim.]
 \vspace{-1.5\parsep}
}
{ \end{proof}}

\theoremstyle{definition}

\theoremstyle{definition}

\def\ppm{\textsc{Second Price Perfect Matching }}
\def\twoppm{\textit{2PPM }}

\def\pm{\textsc{Second Price Matching }}
\def\twopm{\textit{2PM }}

\allowdisplaybreaks[1]

\def\NN{\mathbb{N}}

\newcommand{\set}[1]{\{#1\}}

\newcommand{\rb}[1]{\left(#1\right)}

\pgfplotsset{compat=1.18}

\usepackage{todonotes}

\title{Second Price Matching with Complete Allocation and Degree Constraints}
\author{Rom Pinchasi\footnote{Technion - Israel Institute of Technology, Haifa, Israel, \texttt{rom.pinchasi@gmail.com, room@technion.ac.il}} \quad \quad Neta Singer\footnote{EPFL, Lausanne, Switzerland, \texttt{neta.singer@epfl.ch}} \quad \quad Lukas Vogl\footnote{EPFL, Lausanne, Switzerland, \texttt{lukas.vogl@epfl.ch}. Supported by the Swiss National Science Foundation project 200021-184656 "Randomness in Problem Instances and Randomized Algorithms".} \quad \quad Jiaye Wei\footnote{EPFL, Lausanne, Switzerland, \texttt{jiaye.wei@epfl.ch}}}

\date{}
\begin{document}
\maketitle

\begin{abstract}

We study the \emph{Second Price Matching} problem, introduced by Azar, Birnbaum, Karlin, and Nguyen in 2009. In this problem, a bipartite graph (bidders and goods) is given, and the profit of a matching is the number of matches containing a second unmatched bidder. Maximizing profit is known to be APX-hard and the current best approximation guarantee is $1/2$. APX-hardness even holds when all degrees are bounded by a constant.
In this paper, we investigate the approximability of the problem under regular degree constraints. Our main result is an improved approximation guarantee of $9/10$ for Second Price Matching in $(3,2)$-regular graphs and an exact polynomial-time algorithm for $(d,2)$-regular graphs if $d\geq 4$. Our algorithm and its analysis are based on structural results in non-bipartite matching, in particular the Tutte-Berge formula coupled with novel combinatorial augmentation methods. 

We also introduce a variant of Second Price Matching where all goods have to be matched, which models the setting of expiring goods. We prove that this problem is hard to approximate within a factor better than $(1-1/e)$ and show that the problem can be approximated to a tight $(1-1/e)$ factor by maximizing a submodular function subject to a matroid constraint. We then show that our algorithm also solves this problem exactly on regular degree constrained graphs as above. 
\end{abstract}

\newpage
\section{Introduction}

%


We study \textit{Second Bidder Auctions} in the binary bids setting. Second bidder auction mechanisms are canonical pricing strategies for allocation of goods to bidders, such as for ad allocation used in internet searches. This mechanism is known as the Generalized Second Price mechanism (see e.g. \cite{edelman2007internet}, \cite{varian2007position}). In the second price auction with \textit{binary bids}, there is a set of goods $A$ and a set of bidders $B$. Every bidder $ b \in B$ bids for a subset of goods $S \subseteq A$ which is denoted by a set of edges from $b$ to $S$ of weight $1$. The edges missing from the graph are the $0$ weight bids. 
An auction algorithm must then allocate goods to bidders while maximizing a certain profit function. In the second bidder framework, the profit earned from the set of goods is given by the sum of the matched nodes times the value of their \textit{second} highest bidder.
Thus an algorithm chooses two bidders per good: the bidder with the higher bid is allocated the good and the profit of the allocation is the value of the second bidder. When bids are binary, this setting corresponds to a bipartite input graph $G = (A \cup B, E)$, and a matching $M$ of goods in $A$ such that the profit gained from matching a node $a$ with $b$ is $1$ if and only if there exists a node $b' \in B$ that is a neighbor of $a$ and is not in $M$. This second bidder problem is defined by Azar, Birnbaum, Karlin, and Nguyen \cite{abkn09} as the \textit{Second-Price Matching (2PM)} problem, and they provide lower and upper bounds showing that the problem is APX-hard but that there exists an algorithm that gives a $1/2$ approximation of the maximum profit. In this work, we consider \twopm on degree constrained input graphs. We show that the $1/2$-approximation can be improved for low degree graphs and that for certain regular graphs \twopm can be solved exactly in polynomial time. APX-hardness of \twopm is shown in \cite{abkn09} even for very low degree graphs, thus our work shows that regularity conditions improve the structure of the problem.
\medskip 

\begin{figure}[h!]
    \centering
    \begin{tikzpicture}[every node/.style={draw, circle, minimum size=1.1em, inner sep=1pt},
    A/.style={fill=blue!20}, B/.style={fill=blue!2},
    scale=1.2, >=latex]

    \foreach \i in {1,...,9} {
      \node[B] (b\i) at (-3, 5 - \i) {$b_{\i}$};}
    \foreach \i in {1, ..., 6}{
      \node[B] (a\i) at (-4.5, 3.5 - \i) {$a_{\i}$};
      \draw (b\i) -- (a\i);
    }
    \node[A] (b4) at (-3, 5 - 4) {$b_{4}$};
    \node[A] (b6) at (-3, 5 - 6) {$b_{6}$};
    \node[A] (b7) at (-3, 5 - 7) {$b_{7}$};
    
    \draw (b8) -- (a6);
    \draw (b8) -- (a4);
    \draw (b9) -- (a6);
    \draw (b7) -- (a3);
    \draw (b9) -- (a5);
    \draw (b1) -- (a2);
    \draw (b5) -- (a6);
    \draw (b4) -- (a3);
    \draw (b7) -- (a5);
    \draw (b7) -- (a4);
    \draw (b4) -- (a2);
    \draw (b4) -- (a1);
    \draw (b5) -- (a2);

    \draw[color=red, line width=2.0pt] (b1)--(a1);
    \draw[color=red, line width=2.0pt] (b2)--(a2);
    \draw[color=red, line width=2.0pt] (b3)--(a3);
    \draw[color=red, line width=2.0pt] (b8)--(a4);
    \draw[color=red, line width=2.0pt] (b5)--(a5);
    \draw[color=red, line width=2.0pt] (b9)--(a6);
    \draw[color=blue, line width=2.0pt] (b4)--(a1);
    \draw[color=blue, line width=2.0pt] (b4)--(a2);
    \draw[color=blue, line width=2.0pt] (b4)--(a3);
    \draw[color=blue, line width=2.0pt] (b4)--(a4);
    \draw[color=blue, line width=2.0pt] (b7)--(a5);
    \draw[color=blue, line width=2.0pt] (b6)--(a6);

    \end{tikzpicture}
    
    \caption{Instance of \textit{2PPM} where all edges correspond to weight $1$ bids on the goods of $A$. The perfect matching taken corresponds to the set of red edges, the unmatched nodes are shaded in blue, and the profit gained by the second highest bidder per good is the set of blue edges. As the matching is perfect over the $6$ nodes of $A$, and there exists $6$ blue edges of second price bids, the profit gained here is $6$ which is maximum possible.}
    \label{fig:2ppm}
\end{figure}
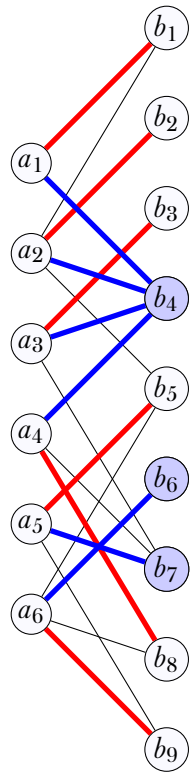

In this work, we also consider a variant of this problem which we call the Second Price \textit{Perfect} Matching (\textit{2PPM}). In the perfect matching variant, the same second price auction with binary bids is studied, with the additional constraint that the matching of the goods in $A$ must be \textit{perfect}. This constraint can be viewed as a welfare constraint such that all goods must get allocated to a bidder even when profit is forfeited. Such a mechanism is consistent with settings where goods expire or must be used up such that perfect allocation is necessary. We formally define the problem statement as follows and give an example instance of this problem in \Cref{fig:2ppm}.

Clearly, a feasible solution to \twoppm is also feasible for \twopm. However, the optima of the two problems may be far apart for identical inputs due to the requirement of the underlying matching $M$ being perfect even when some matches do not increase the profit. Interestingly, the perfect matching variant poses more difficulties and therefore admits a stronger approximation lower bound of $(1- 1/e)$ which we prove in this paper. However, the problem admits a natural reformulation as a submodular maximization problem over a matroid constraint such that canonical algorithms give a $(1-1/e)$-approximation guarantee beating the $1/2$-approximation guarantee of \cite{abkn09} for \twopm. Thus for the perfect matching variant, the approximation guarantee of the problem is tight without added assumptions on the input. Interestingly, this shows that \twoppm behaves more similarly than \twopm to the more general submodular welfare maximization (SWM) and maximum budgeted allocation (MBA) problems, which are first price analogs of the general allocation framework. This is as these problems admit natural $(1-1/e)$-approximation algorithms as well (see e.g. \cite{vondrak2008optimal}, \cite{andelman2004auctions}) and for SMW, hardness results show that $(1-1/e)$ is best possible in the value oracle model \cite{khot2005inapproximability}. The perfect matching variant of second price matching thus behaves more similarly to SMW and MBA on the complexity side than the original \twopm formulation. It remains an interesting open question whether the perfect matching variant can be leveraged to provide tighter guarantees for the original \twopm problem where the gap between APX-hardness and the $1/2$-approximation algorithm remains large. 

\begin{center}
    \begin{algorithm}[H]\label{prob:BN}
    \caption{\;\textsc{Second Price Matching $\left(2PM\right)$}}
    \begin{algorithmic}
        \Require A bipartite graph $G=(A \cup B, E)$. 
        \Ensure Subsets $S \subseteq B$, $W \subseteq A$ which maximize the number of elements of $W$ that have neighbors in $S$ while maintaining a matching between $W$ and $B \setminus S$.
    \end{algorithmic}
    \end{algorithm}
\end{center}

\begin{center}
    \begin{algorithm}[H]\label{prob:BN}
    \caption{\;\textsc{Second Price Perfect Matching $\left(2PPM\right)$}}
    \begin{algorithmic}
        \Require A bipartite graph $G=(A \cup B, E)$ where $|A| <|B|$ and there exists an $A$-perfect matching in $G$. 
        \Ensure A subset $S \subseteq B$ which maximizes the number of elements in $A$ that have neighbors in $S$ while maintaining an $A$-perfect matching between $A$ and $B\setminus S$.
    \end{algorithmic}
    \end{algorithm}
\end{center}

Problem 2 is the analog of Problem 1 with a perfect matching constraint. Problem 1 is a reformulation of \twopm, as the set $S$ corresponds to a set of second highest bidders and the number of items in $A$ with a neighbor in $S$ is the profit earned from allocating all goods in $A$ to bidders in $B\setminus S$ perfectly. Indeed the profit earned under binary bids is simply the number of goods with a nonzero edge to $S$.


\subsection{Our Contributions}
As our \emph{main contribution}, we show that \twopm can be solved in polynomial time for $(d, 2)$-regular graphs when $d \geq 4$ and we give an improved approximation algorithm of $(9/10)$ for $(3, 2)$-regular graphs. 
This shows that right-degree constrained binary bid auctions are tractable to compute optimally in contrast to the general bidding mechanism. This may occur for example in a mechanism where bidders are restricted to $2$ bids each or where bidder budgets do not allow for making several bids. 

These results are based on finding a \emph{non-bipartite} maximum matching in an auxiliary graph followed by a combinatorial augmentation algorithm, that yields a solution to $(d, 2)$-regular \twopm, optimal for $d\geq 4$ and near-optimal for $d=3$. The Tutte-Berge formula is applied to get an approximation guarantee of the $(3, 2)$-regular case.

\bigskip
\noindent 
Furthermore, we introduce a variant of the matching problem \textit{2PPM}, where all goods need to be matched. We settle the approximability of this problem on both the hardness and the algorithmic side. In particular, we give inapproximability lower bounds for \textit{2PPM} in its fullest generality. We also show that if right degrees are increased to up to $4$ bids per bidder, \textit{2PPM} becomes APX-hard even when left degrees are small. However, we show that our algorithm for $(d, 2)$-regular graphs also solves \twoppm optimally in polynomial time for any $d\geq 3$. Our main hardness result is formulated as follows.

\begin{restatable}{theorem}{inapx}\label{thm: 1-1/e-inapprox}
    For any $\varepsilon>0$, \twoppm cannot be approximated within a ratio of $(1-1/e+\varepsilon)$ in polynomial-time unless P = NP.
\end{restatable}

This lower bound is not only stronger than the APX-hardness of \twopm, but is in fact tight as \textit{2PPM} can be reformulated as a coverage maximization problem over a single matroid constraint, which is known to have a $(1-1/e)$ approximation algorithm. Thus in the general case, \Cref{thm: 1-1/e-inapprox} shows that coverage maximization algorithms are best possible for \textit{2PPM}.

The following theorem states that even when the degree of the input graph is bounded by low degrees, the problem \twoppm remains APX-hard. 

\begin{restatable}{theorem}{apxhard}\label{thm: APX-hardness}
    It is NP-hard to approximate \twoppm with $\deg(a) \in \{2,3\}$ for every $a\in A$ and $\deg(b) \in \{1,2,4\}$ for every $b\in B$, to within a factor of 293/297.
\end{restatable}

\Cref{thm: APX-hardness} shows hardness even for special inputs with very low degree. However, we will show that for one sided degree bounded inputs the problem can be solved exactly in polynomial time. In particular, given any regular bipartite graph as input where either $A$ or $B$ is $2$-regular, \textit{2PPM} becomes polynomial time solvable via reductions to a general graph matching problem. This reduction is identically used to provide exact algorithms for \twopm. Our main algorithmic results for \twopm and \twoppm are stated as follows.




\begin{restatable}{theorem}{dtworegular}\label{thm: d-2-regular}
    On input $G = (A \cup B, E)$, where $\deg(a) = d$ for $a \in A$ ($d\geq 3$) and $\deg(b) = 2$ for $b \in B$, the following holds:
    \begin{enumerate}[label = (\arabic*)]
        \item If $d=3$, there exists a polynomial-time algorithm which gives a $9/10$-approximation of \twopm and solves \twoppm exactly.
        \item If $d\geq 4$, there exists a polynomial-time algorithm which solves \twopm and \twoppm exactly.
    \end{enumerate}
\end{restatable}


In \Cref{sec: deg a is 2}, we also show that for inputs of the form $G = (A \cup B, E)$, $\deg(a) = 2$ for every $a \in A$, \ppm can also be solved exactly in polynomial time. Together with \Cref{thm: d-2-regular}, this shows that even though the low degree case of \textit{2PPM} is APX-hard, when at least one side of the bipartite graph is 2-regular the problem becomes tractable.

\subsection{Related Works}
\twopm and \textit{2PPM} are natural combinatorial variants of the second price auction mechanism. The second price auction falls more generally into the context of combinatorial allocations for which there is a rich body of literature. Combinatorial allocation problems are generally given by a set of goods which must be allocated to competing bidders while maximizing some profit or utility of the allocation. Much of the prior work in this field is focused on first price mechanisms where profit is exactly the value of the allocated bid. One common example is the first price ad auction, which is shown to be NP-hard but for which there exists a $4/3$ approximation algorithm \cite{chakrabarty2010approximability}, \cite{srinivasan2008budgeted} which was subsequently improved to $4/3+c$ for an absolute constant $c$ \cite{kalaitzis2016improved}. Prior to these results, weaker approximation guarantees for first price ad auctions were shown through a series of works \cite{garg2001approximation}, \cite{lehmann2001combinatorial}, \cite{andelman2004auctions}, \cite{azar2008improved}.

In \cite{mehta2007adwords}, an improved approximation guarantee for first price ad auctions is shown when bidder budgets far exceed the value of their bids. In our work, we show a converse condition where bidders only bid on two items each thus modeling a low budget of $2$ in comparison to the bid values which are uniformly $1$. In this second price mechanism, it is therefore useful for bidders to have small bid allowances as opposed to large ones.

In \cite{buchbinder2007online}, bounded degree graphs are also considered, where similarly to our setting, few bidders will bid for the same good. They consider a special case where goods have set prices and bidders only decide whether to bid for the good or not. In this context, they give a $1 - \frac{d-1}{d(1 + \frac{1}{d+1})^{d-1}}$ approximation algorithm for degree $d$ bounded graphs. 


To the best of our knowledge, \twopm is the first study of the ad allocation problem under the second bidder pricing scheme and assuming binary bids. Other variants of \textit{2PM} have been considered since then. Fernandes and Schouery generalize \textit{2PM} to allow budgets per bidder that are larger than $1$ \cite{fernandes2014second}. The online version of first and second price ad auctions has also been studied extensively, see e.g. \cite{buchbinder2007online}, \cite{mehta2007adwords}, \cite{aggarwal2011online}, \cite{abkn09} among many other works. We refer interested readers to \cite{mehta2013online} for a survey on the online ad allocation problem.

\section{Algorithms for Second Price Matching and Second Price Perfect Matching}
In this section we present our main algorithm for the \textsc{Second Price Matching} problem and its special cases.
We show that our polynomial time algorithm solves \twopm and \twoppm to optimality for the degree constrained problem, when bidders bid on $2$ items each. In this setting, the input graph is assumed to have regular vertex degrees and by using this structure, we efficiently construct an optimal solution to the problem.

We also show that \ppm admits a formulation as a maximization problem over a monotone submodular function subject to a matroid constraint. Without leveraging any specific properties of \ppm, the maximum of the submodular function can be approximated by a factor $(1-1/e)$ by a canonical tight submodular maximization approximation algorithm \cite{calinescu2007maximizing}. 


\subsection{Algorithms for $(d,2)$-regular 2PM and 2PPM}\label{sec:d-2-regular}
We turn to focusing on the special case of \twopm and \twoppm where in the input graph $G=(A\cup B, E)$, $A$ is $d$-regular and $B$ is 2-regular, which we call the $(d,2)$-regular \twopm and \twoppm respectively.
Let $|A|=n$ and $|B|=m$. Note that $d\geq 3$ since $n<m$. 
Our main theorem is that there exists a polynomial-time algorithm $\mathcal{A}_1$ which gives a 9/10-approximation of $(3,2)$-regular \twopm and solves $(3,2)$-regular \twoppm exactly, and for $d\geq 4$ there exists a polynomial-time algorithm $\mathcal{A}_2$ which solves $(d,2)$-regular \twopm and $(d,2)$-regular \twoppm exactly.
Our result largely improves over the 1/2-approximation for \twopm given in \cite{abkn09} and the $(1-1/e)$-approximation for \twoppm given by the submodular maximization framework in (which we show in Section \ref{sec:submodular}). 

\dtworegular*

To prove Theorem \ref{thm: d-2-regular}, we will separate the cases where $d=3$ and $d\geq 4$ and prove the statements by leveraging the degree constraints on the instance $G$.
As we will see, the case $d=3$ is considerably different and more difficult than the case $d \geq 4$.
The following construction is important for both cases and will be used throughout this section.

\begin{figure}[htbp]
\centering
\begin{subfigure}[b]{0.45\textwidth}
\centering
\begin{tikzpicture}[
    node distance=1.0cm,
    vertex/.style={circle, draw, minimum size=0.5cm},
    edge/.style={draw}]
    
    \node[vertex] (A1) at (0,2.5) {$a_1$};
    \node[vertex] (A2) at (0,1.5) {$a_2$};
    \node[vertex] (A3) at (0,0.5) {$a_3$};
    
    \node[vertex] (B1) at (2,3) {$b_1$};
    \node[vertex] (B2) at (2,2) {$b_2$};
    \node[vertex] (B3) at (2,1) {$b_3$};
    \node[vertex] (B4) at (2,0) {$b_4$};
    
    \draw[edge] (B1) -- (A1);
    \draw[edge] (B1) -- (A2);
    \draw[edge] (B2) -- (A1);
    \draw[edge] (B2) -- (A2);
    \draw[edge] (B3) -- (A1);
    \draw[edge] (B3) -- (A3);
    \draw[edge] (B4) -- (A2);
    \draw[edge] (B4) -- (A3);
\end{tikzpicture}
\caption{The original graph $G$ where $B$ is 2-regular}
\end{subfigure}
\hfill
\begin{subfigure}[b]{0.45\textwidth}
\centering
\begin{tikzpicture}[
    node distance=1.5cm,
    vertex/.style={circle, draw, minimum size=0.7cm},
    edge/.style={draw}]
    
    \node[vertex] (A1) at (1.5,3) {$a_1$};
    \node[vertex] (A2) at (0,0) {$a_2$};
    \node[vertex] (A3) at (3,0) {$a_3$};
    
    \draw[edge] (A1) to[bend left=20] node[below right] {$e_{b_2}$} (A2);
    \draw[edge] (A1) to[bend right=20] node[above left] {$e_{b_1}$} (A2);
    \draw[edge] (A1) -- node[above right] {$e_{b_3}$} (A3);
    \draw[edge] (A2) -- node[below] {$e_{b_4}$} (A3);
\end{tikzpicture}
\caption{The auxiliary graph $G^\prime$}
\end{subfigure}
\caption{The construction of the auxiliary graph $G^\prime$}
\label{fig:construction-G-prime}
\end{figure}
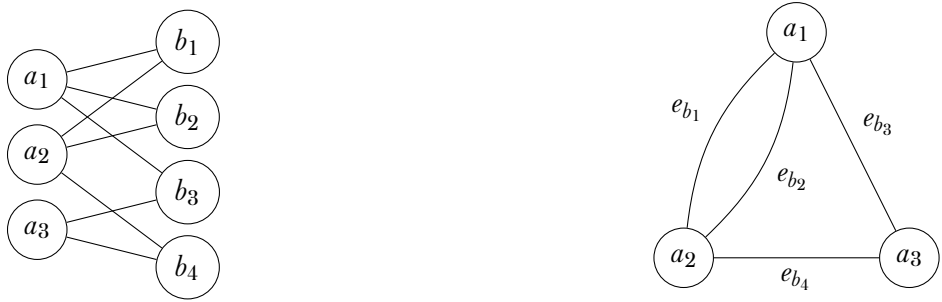

\paragraph{Construction of the auxiliary graph $G^\prime$}
For any instance $G=(A\cup B, E)$ of both \twopm and \twoppm where $B$ is 2-regular\footnote{The construction of $G^\prime$ requires no degree constraints on $A$.}, we construct an auxiliary graph $G^\prime = (V^\prime, E^\prime)$ with $V^\prime = A$ where two vertices $a_1$ and $a_2$ are connected by an edge $e_b \in E^\prime$ if $a_1, a_2$ share a common neighbor $b$ in $G$. 
Let $\varphi: B \rightarrow E^\prime$ be the bijection $b \mapsto e_b$.
Notice that the auxiliary graph $G^\prime$ may have multiple-edges, since two vertices of $A$ may share more than one common neighbor. We illustrate the construction in Figure \ref{fig:construction-G-prime}.
If $A$ is $d$-regular, then the auxiliary graph $G^\prime$ is also $d$-regular. 

Having defined $G'$, a maximum-size matching $M^\prime \subseteq E^\prime$ in $G^\prime$ can be found in polynomial-time using the blossom algorithm. We refer the reader to \cite{sch03} for a detailed description of the blossom algorithm. The size of this maximum matching is traditionally denote by $\nu(G')$.
The edges of $M^\prime$ correspond to the subset $B':=\varphi^{-1}(M^\prime)$ of $B$. Notice that $|B'|=|M^\prime|=\nu(G^\prime)$. We also have $|N(B')| = 2|B'|$. 
By removing the subset $B'$ and all the edges incident to $B'$ from $G$, we get an induced bipartite subgraph $H$ of $G$ whose two parts are $A$ and $B\setminus B'$. 

\subsubsection{The Case $d=3$}
We start by proving Theorem \ref{thm: d-2-regular} for $(3,2)$-regular \twoppm, which we state here as a separate theorem.

\begin{theorem}\label{thm:3-2-regular}
    The $(3,2)$-regular \twoppm can be solved exactly in polynomial-time. 
    Furthermore, the optimal value $\text{OPT}_{2PPM}$ is equal to $n/2 + \nu(G^\prime)$ where $G^\prime$ is the auxiliary graph of $G$ constructed above and $\nu(G^\prime)$ is the size of maximum matching in $G^\prime$. 
\end{theorem}

\begin{proof}
    When $d=3$, we have $m = 3n/2$ and the subset $S\subseteq B$ to be chosen in the \textsc{2PPM} problem is of size (at most) $n/2$. 
    Recall the construction of the auxiliary graph $G^\prime$, the  maximum matching $M^\prime$ in $G^\prime$, and its corresponding subset $B':=\varphi^{-1}(M^\prime) \subseteq B$ of size $|B'|=|M^\prime|=\nu(G^\prime) \leq n/2$. 
    We recall also that $H$ is the subgraph of $G$ induced by $A \cup (B\setminus B')$. For the graph $H$ and a vertex $a \in A$ we have
    \[
    \deg(a) = 
    \begin{cases}
        3, & \text{if}\;a\in A\setminus N(B'),\\
        2, & \text{if}\;a\in N(B'),
    \end{cases}
    \quad
    \text{and}
    \quad
    \deg(b) = 2, \;\text{for every}\; b\in B\setminus B'.
    \]

    To prove \Cref{thm:3-2-regular} we will find in polynomial-time a set $S \subseteq B$ by augmenting $B^\prime$, such that $|N(S)| = n/2 + \nu(G^\prime)$ and there is an $A$-perfect matching in the bipartite induced subgraph of $G$ between $A$ and $B\setminus S$. 
    We will also show that this value is optimal and cannot be improved.

        We start by partitioning $H$ into (vertex-disjoint even) cycles $C_1, \dots, C_\ell$ and a forest $F$. The cycles $C_{i}$ are necessarily vertex-disjoint because the maximum degree of a vertex in $H$ is equal to $3$. 
        Notice that all the leaves of $F$ are from $A\setminus N(B')$ because all other vertices of $H$ have an even degree (more precisely, $2$) in $H$. Notice moreover that a leaf $x$ of $F$ must belong to exactly one of the cycles $C_i$ because originally $x$ had degree $3$ in $H$.
        
        Next we remove all the leaves of $F$ to get a smaller forest $F^\prime$. The leaves of $F^\prime$, if there are any, all belong to $B\setminus B'$. Observe that every leaf $x$ of $F^\prime$ must have at least one neighbor in $H$ in the set $A\setminus N(B')$. Indeed, this neighbor is a leaf of $F$ that was removed when we obtained $F'$. 
        In fact, any leaf $x$ of $F^\prime$ has in $H$ \textbf{precisely} one neighbor in $A\setminus N(B')$, otherwise we could have matched both of its neighbors in $A\setminus N(B')$ and add it as an edge to $M^\prime$ to get a larger matching in $G^\prime$, contradicting the maximality of $M^\prime$. We conclude that the neighbor in $F'$ of any leaf $b$ (necessarily in $B \setminus B'$) must belong to $N(B')$.
    
        We generate a collection of vertex-disjoint paths in $F'$ by iteratively doing the following: Find a path $P$ between any two leaves of $F^\prime$, then remove the vertices of the path $P$ and all of their incident edges from $F^\prime$ to obtain a smaller forest $F''$, that we consider as $F'$ for the next iteration. 

In the next two claims we establish simple properties of $F''$ that will be crucial later.

        \begin{claim}\label{claim:path}
        After removing the path $P$ from $F'$ the degrees of the remaining vertices of $A$ in $F''$ is not changed.
        Moreover, all the leaves, if there are any, in $F''$ must still belong to
        $B \setminus B'$. 
        \end{claim}

        \begin{cproof}
        Let $x \in A$ be a vertex in $F'$ that does not belong to $P$. The neighbors of $x$ are all from $B$. However, the vertices of $B$ in the path $P$ do not have in $F'$ any neighbor that is not in $P$. This is because their degrees in $F'$ are either $1$ or $2$. We conclude that $x$ cannot have any neighbor in $P$. Consequently, removing the vertices of $P$ from $F'$ does not affect the degree of $x$.

        The second part of the claim now follows because all the leaves in $F'$ are from $B$. Removing the vertices of $P$ from $F'$ does not change the degrees of the remaining vertices in $A$ and in particular they cannot become leaves
        in $F''$.
        \end{cproof}

        \begin{claim}\label{claim:leaf}
        Every leaf $b$ in the remaining forest $F''$ must have in the graph $H$ one neighbor in $A \setminus N(B')$ and one neighbor in $N(B')$. The neighbor of $b$ in $F''$ is necessarily the one in $N(B')$.
        \end{claim}
        
        \begin{cproof} The proof is by induction on the iterative construction.
        We know already that the leaf $b$ must belong to $B \setminus B'$.
        If $b$ is a leaf already in $F'$, then the claim is true by the induction hypothesis (we have already seen the basis of induction when we moved from $F$ to $F'$).
        Otherwise, $b$ is not a leaf in $F'$ and it becomes a leaf
        in $F''$ after we removed the path $P$ from $F'$. We conclude that $b$ has in $F'$ a neighbor that is a vertex
        in the path $P$. Because $b \in B$ the neighbor of $b$ in $P$ must be a vertex $a \in A$. The degree of $a$ in $P$ is equal lto $2$ and therefore the degree of $a$ in $H$ is equal to $3$. We conclude that $a$ is in $A \setminus N(B')$. This shows that $b$ has one neighbor in $A \setminus N(B')$ and this neighbor does not belong to $F''$.
        It cannot be that $b$ has two neighbors in $A \setminus N(B')$ because that would contradict the maximality of $M'$ as we could add the two neighbors of $b$ to the matching $M'$. We conclude that $b$ must have one neighbor in $N(B')$
        and this must be the neighbor of $b$ in $F''$.
        \end{cproof}

        We continue iteratively removing paths connecting two leaves from the remaining forest $F''$, which we now consider as our new $F'$.
        
        We stop when there are no edges in $F^\prime$. This way we generate a collection of paths $P_1, \dots, P_k$, where each path starts and ends at vertices of $B\setminus B'$. 
        because of Claim \ref{claim:leaf}, we know that in every such path $P_{j}$ each of the two end vertices is connected to a vertex from $N(B')$.

In the following two claims we establish more properties of the cycles and the paths that we have constructed.

\begin{claim}\label{claim:one_a}
        Every $a \in A$ belongs either to exactly one of the cycles $C_i$, or to exactly one of the paths $P_j$.
        \end{claim}

        \begin{cproof}
        Consider a vertex $a \in A$ that does not belong to any cycle $C_{i}$. Then the degree of $a$ in 
        $F$ is equal to the degree of $a$ in $H$ that is either $2$ or $3$. If $a$ does not belong to any of the paths $P_{j}$, then by Claim \ref{claim:path}, the degree of $a$ will never change. This is impossible because when we stop there are no edges in the remaining graph $F''$.
        This shows that $a$ must belong to a cycle $C_{i}$ or to a path $P_{j}$. The degree of $a$ in a cycle $C_{i}$ or in a path $P_{j}$ to which it belongs must be equal to $2$ because every path $P_{j}$ starts and ends at vertices of $B \setminus B'$. Because the degree of $a$ in $H$ is at most $3$ we conclude that $a$ cannot belong to more than one cycle $C_{i}$ or a path $P_{j}$.
        \end{cproof}

\begin{claim}\label{claim:one_b}
        Every vertex of $B \setminus B'$ either belongs to exactly one of the cycles $C_i$, or it is an internal vertex of exactly one of the paths $P_j$, or it was a leaf in $F'$ at some iteration of our construction.
        \end{claim}

        \begin{cproof}
        Consider a vertex $b \in B \setminus B'$ that does not belong to any cycle $C_{i}$ and is not an internal vertex in any path $P_{j}$. Then the degree of $b$ in 
        $F$ is equal to the degree of $b$ in $H$ that is $2$. Because when we stop there are no edges in $F'$, consider the first time that an edge incident to $b$ was removed. It cannot be when we moved from $F$ to $F'$ because $b$ is not a leaf in $F$. 
        The degree of $b$ in $F'$ can be reduced from $2$ only when $b$ is an internal vertex of a path $P_{j}$ that we remove from $F'$, or when $b$ is a neighbor of a vertex in a path $P_{j}$ that we remove from $F'$. In the first case we are done. In the later case we notice that $b$ cannot be a neighbor of two vertices from a path $P_{j}$ that we remove from $F'$ because $F'$ is a forest. We conclude that after we remove $P_{j}$ from $F'$ the vertex $b$ becomes a leaf, as desired.
        \end{cproof}

        Having constructed 
        the cycles $C_{1}, \ldots, C_{\ell}$ and the paths
        $P_{1}, \ldots, P_{k}$, and having established some of their properties, we construct an $A$-perfect matching $M$ in $H$. Then the desired set $S$ will be the set of vertices in $B$ that do not participate in $M$.
        
        In order to construct $M$, for every $i=1, \ldots, \ell$ we take a perfect matching in $C_{i}$ between vertices in $A$ to vertices in $B \setminus B'$. There are two possibilities for such a matching within each $C_{i}$ and we arbitrarily choose one of them. 
        Similarly, for every $j=1 \ldots, k$ 
        we consider the path $P_{j}$ and ignore one of the end vertices of it. Then we take a perfect matching in $P_{j}$ for the remaining vertices of $P_{j}$.

We define $M$ to be the union of all the matchings that we constructed within all the cycles $C_{i}$ and all the paths $P_{j}$.
By Claim \ref{claim:one_a}, every vertex in $A$ belongs either to precisely one cycle $C_{i}$, or to precisely one path $P_{j}$. 
By Claim \ref{claim:one_b}, no vertex in $B$ may belong to more than one cycle $C_{i}$ or path $P_{j}$.
We conclude that $M$ is indeed an $A$-perfect matching in $H$.
        
        Let $S$ be the union of $B'$ and the set of those vertices of $B \setminus B'$ that are not in $M$.

\begin{claim}\label{claim:finish}
In order to show that $N(S)=n/2+\nu(G^\prime)$ it is enough 
        to show that no vertex in $A \setminus N(B')$ has two (or more) neighbors in $S \setminus B'$.
        \end{claim}

        \begin{cproof}

The cardinality of 
        $S \setminus B'$ is equal to $\frac{3}{2}n-n-\nu(G')=n/2-\nu(G')$ because there are $\frac{3}{2}n$ vertices in $B$ while $n$ of them participate in $M$ and another $\nu(G')$ of them are in $B'$.

Consider now any vertex $b$ in $S \setminus B'$. 
By Claim \ref{claim:one_b} and by our construction, the vertex $b$ must have been leaf in $F'$ at some iteration of our construction. By Claim \ref{claim:leaf}, $b$ must have a neighbor in $N(B')$ and one neighbor in $A \setminus N(B')$.         
        If we can show that these neighbors in
        $A \setminus N(B')$ of vertices in $S \setminus B'$ are all distinct, then we may conclude that 
        
        $$
        |N(S)|=|N(B')|+|S \setminus B'|=
        2\nu(G')+n/2-\nu(G')=n/2+\nu(G'),
        $$
        as desired.
\end{cproof}

In view of Claim \ref{claim:finish}, in order to show $N(S)=n/2+\nu(G^\prime)$ it is left to show that no vertex in $A \setminus N(B')$ can have
two neighbors in $S \setminus B'$. 

Consider a vertex $a \in A \setminus N(B')$. If $a$ belongs to a cycle $C_{i}$, then its two neighbors in $C_{i}$ participate in $M$
and therefore are not in $S$. Because the degree of $a$ in $H$ is $3$, the vertex $a$ can have at most one neighbor in $S$.
If $a$ does not belong to a cycle $C_{i}$, then it must belong to a path $P_{j}$. Because $a \notin N(B')$, then by Claim \ref{claim:leaf}, the two neighbors of $a$ in $P_{j}$ are both internal vertices of $P_{j}$ and none of them can be an end vertex of $P_{j}$.
Therefore, by our construction, the two neighbors of $a$ in $P_{j}$ participate in the matching $M$. Once again we conclude that $a$ can have at most one neighbor in $S$. This concludes the argument showing that $N(S)=n/2+\nu(G^\prime)$.

We conclude the proof of \Cref{thm:3-2-regular} by showing that the feasible solution $S$ constructed in the claim above, is also optimal.
    To prove this, take any feasible solution $T\subseteq B$ of the \textsc{2PPM} on $G$ where $|T|=n/2$. There exists a maximal subset $T^\prime \subseteq T$ that corresponds to a matching in the auxiliary graph $G^\prime$, i.e., the vertices in $T^\prime$ have pairwise-disjoint neighbors in $G$. 
    Every vertex $b$ in $T\setminus T^\prime$ has at most one neighbor that is not in $N(T')$, otherwise we can add $b$ to $T'$ thus contradicting the maximality of $T^\prime$. Therefore, we have $|N(T)| \leq 1\cdot (|T|-|T^\prime|)+2\cdot |T^\prime|=|T|+|T^\prime| \leq n/2 + \nu(G^\prime) = |N(S)|$.
\end{proof}

\medskip

\Cref{thm:3-2-regular} shows that the optimal value $\text{OPT}_{2PPM}$ of the (3,2)-regular \textsc{2PPM} problem
depends only on the size of the maximum matching in the auxiliary graph $G'$ and the number $n$ of vertices in $A$. By providing good, and hopefully tight, lower bounds for $\nu(G')$, we can bound from below the optimal value $\text{OPT}_{2PPM}$ of the (3,2)-regular \textsc{2PPM} problem only in terms of $n$ and regardless of the input graph $G$. This is then used to obtain a $9/10$-approximation guarantee for $(3,2)$-regular \twopm. This is done in the following proposition.

\begin{proposition}\label{prop:lower-bound-(3,2)-2PPM}
    The optimal value satisfies $\text{OPT}_{2PPM}\geq 9n/10$ where $n=|A|$ for any instance of (3,2)-regular \twoppm.
    This lower bound of $\text{OPT}_{2PPM}$ is tight.
\end{proposition}

\begin{proof}
First we prove the following claim that gives a lower bound on the size of a maximum matching in $3$-regular graphs with possibly multiple edges. \footnote{We thank StackExchange user Matthias for asking a similar question for simple graph \cite{StackExchange}, and user Misha Lavrov for providing a solution.}
\begin{claim}\label{claim:3-reg-max-match}
    For any 3-regular graph $G=(V,E)$ possibly with multiple-edges, we have $\nu(G)\geq 2|V|/5$.
\end{claim}
\begin{cproof}
    Recall the Tutte-Berge formula (see \cite{sch03}) by which the maximum size of a matching in a (simple) graph $G$ is equal to
    \[
    \frac{1}{2}\min_{U \subseteq V}\rb{|U| - \text{odd}(G-U) + |V|}
    \]
    where $\text{odd}(G-U)$ is the number of odd connected components of $G-U$. 
    Consider now any $3$-regular graph $G=(V,E)$ and let $|V|=n$.
    Fix $U \subseteq V$ and let $x$ be the number of connected components of $G-U$ that consist of a single vertex. Let $y$ be the number of odd connected components of $G-U$ that have at least $3$ vertices. Every connected component of $G-U$ that is a single vertex has $3$ edges going to $U$; Every odd connected component of $G-U$ of size at least $3$ has at least $1$ edge going to $U$. Indeed, otherwise this connected component is by itself a $3$-regular graph with an odd number of vertices which is impossible. Evaluating from above and from below the number $E(\text{odd components of}\;G-U, U)$ of edges from the odd components of $G-U$ to $U$, we obtain the following inequality
    
    \begin{equation}\label{eqn:counting_edges}
        3x + y \leq |E(\text{odd components of}\;G-U, U)| \leq 3|U|.
    \end{equation}

    Evaluating from above and from below the number $V(\text{odd components of}\;G-U)$ of vertices in all the odd components of $G-U$ together, we obtain the inequality
    
    \begin{equation}\label{eqn:counting_vertices}
        x + 3y \leq |V(\text{odd components of}\;G-U)| \leq n - |U|.
    \end{equation}
    By summing up the inequalities (\ref{eqn:counting_edges}) and (\ref{eqn:counting_vertices}) with weights $2/5$ and $1/5$, respectively, we get
    \begin{equation}
        \frac{1}{5}n + |U| \geq \frac{7}{5}x + y \geq x+y \quad \Rightarrow \quad |U| - (x+y) \geq -\frac{1}{5}n.
    \end{equation}
    
    Becase $x+y=\text{odd}(G-U)$, this implies the desired lower bound on the size of a maximum matching in $G$:
    \begin{equation}
        \frac{1}{2}\min_{U \subseteq V}\rb{|U| - \text{odd}(G-U) + n} \geq \frac{2}{5}n.
    \end{equation}
\end{cproof}

We can now finish the proof of Proposition \ref{prop:lower-bound-(3,2)-2PPM}. By \Cref{thm:3-2-regular}, for $(3,2)$-regular \textsc{2PPM} problem we have $\text{OPT}_{2PPM} = n/2+\nu(G^\prime)$. By Claim \ref{claim:3-reg-max-match}, we have $\text{OPT}_{2PPM} = n/2+\nu(G^\prime) \geq n/2+2n/5=9n/10$.

It is left to show the tightness of the lower bound of $\text{OPT}_{2PPM}$ in Proposition \ref{prop:lower-bound-(3,2)-2PPM}.
To this end it is enough to show that the lower bound $\nu(G^\prime) \geq 2n/5$ in Claim \ref{claim:3-reg-max-match} is tight for $3$-regular graph $G^\prime$ with $n$ vertices. 
We consider the graph shown in Figure \ref{fig:3-2-tightness} where the graph contains $n=10$ vertices and the size of maximum matching equals to $4$. One can generate larger examples by taking 
several disjoint copies of this small graph of $10$ vertices.

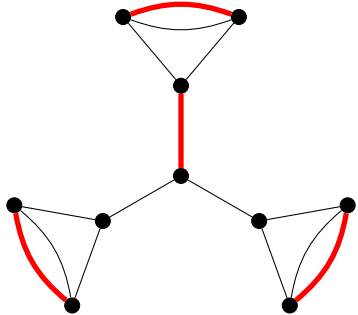
\begin{figure}[h]
    \centering
    \begin{tikzpicture}[scale=1.5, every node/.style={circle, draw, fill=black, inner sep=2pt}]
        \node (C) at (0,0) {};
        \node (A) at (90:0.8) {};    
        \node (B) at (210:0.8) {};   
        \node (D) at (330:0.8) {};   
    
        \foreach \v in {A, B, D} {
            \draw (C) -- (\v);
        }
        \draw[color=red, line width=2.0pt] (C) -- (A);
    
        \node (A1) at (90+20:1.5) {};
        \node (A2) at (90-20:1.5) {};
        \node (B1) at (210+20:1.5) {};
        \node (B2) at (210-20:1.5) {};
        \node (D1) at (330+20:1.5) {};
        \node (D2) at (330-20:1.5) {};
    
        \draw (A1) -- (A) -- (A2);
        \draw (B1) -- (B) -- (B2);
        \draw (D1) -- (D) -- (D2);
    
        \draw[color=red, line width=2.0pt] (A1) to[out=20, in=160] (A2);
        \draw (A1) to[out=-20, in=-160] (A2);
        \draw[color=red, line width=2.0pt] (B1) to[out=140, in=280] (B2);
        \draw (B1) to[out=100, in=320] (B2);    
        \draw[color=red, line width=2.0pt] (D1) to[out=260, in=400] (D2);
        \draw (D1) to[out=220, in=440] (D2);
    \end{tikzpicture}
    \caption{A tight example of 3-regular graph $G$ (with multiple edges) with $\nu(G)=2|V(G)|/5$. One of the maximum matchings in $G$ is colored in red.}
    \label{fig:3-2-tightness}
\end{figure}
\end{proof}

We have shown that through the auxiliary graph $G'$, we are able to solve for an optimal solution to \twoppm on $(3,2)$-regular graphs. We may now conclude the first part of \Cref{thm: d-2-regular} by showing that this solution is also a $9/10$-approximation of the optimum for \twopm. 

\begin{theorem}\label{thm:3-2-regular 2pm}
    There exists an algorithm that outputs a $9/10$-approximation of the optimal solution of $(3,2)$-regular \twopm.
\end{theorem}

\begin{proof}
By \Cref{thm:3-2-regular}, we can find in polynomial time a solution $S \subseteq B$ that is optimal for \twoppm. Notice that by definition, this solution is also feasible for \twopm by taking $S \subseteq B$ and letting $W=A$. 

To analyze the quality of this feasible solution, by Proposition \ref{prop:lower-bound-(3,2)-2PPM}, we have that
\[
|N(S) \cap W| = |N(S)| = \text{OPT}_{2PPM} \geq \frac{9}{10}n \geq \frac{9}{10}\text{OPT}_{2PM}
\]
where the last inequality follows from $\text{OPT}_{2PM} \leq n$ for any input graph by definition.
\end{proof}

\Cref{thm:3-2-regular} and \Cref{thm:3-2-regular 2pm} together conclude the proof of the first part of Theorem \ref{thm: d-2-regular} in the case that $d=3$.


\subsubsection{The Case $d \geq 4$}
We now show the second part of \Cref{thm: d-2-regular}, for inputs $G=(A\cup B, E)$ where the vertices in $A$ each have degree $d\geq 4$, then both $(d,2)$-regular \twopm and \twoppm can be solved exactly in polynomial-time.

\begin{theorem}\label{thm:d-2-regular}
    The $(d,2)$-regular \twopm and \twoppm with $d\geq 4$ can be solved exactly in polynomial-time. 
    Furthermore, the optimal values $\text{OPT}_{2PM}$ and $\text{OPT}_{2PPM}$ are both equal to $n=|A|$.
\end{theorem}

\begin{proof}
    Recall that for any instance $G$ of $(d,2)$-regular \twopm and \twoppm we construct an auxiliary graph $G^\prime$, a maximum-size matching $M^\prime$ in $G^\prime$ and its corresponding subset $B':=\varphi^{-1}(M^\prime) \subseteq B$, and an induced bipartite subgraph $H$ of $G$ with parts $A$ and $B\setminus B'$.
    Because of the maximality of $M^\prime$, we have the following.
    
    \begin{observation}\label{observation:1}
    The vertices of $A\setminus N(B')$ have pairwise-disjoint neighbors in $H$, i.e., for any $a_1, a_2 \in A\setminus N(B'), a_1\neq a_2$, we have $N(a_1) \cap N(a_2) = \emptyset$. In other words, every vertex $b \in B\setminus B'$ has at most one neighbor in $A\setminus N(B')$.
    \end{observation}
    
    We use the following construction for augmenting the subset $B' \subseteq B$ to get an optimal solution of \twopm and \twoppm.    
    
    \paragraph{Construction of the auxiliary bipartite graph $G^{\prime\prime}$}
    
    We construct another auxiliary graph $G^{\prime\prime}$ based on $G$ and the subset $B'$. The graph $G^{\prime\prime}$ is a bipartite graph with parts $A\setminus N(B')$ and $N(B')$. We connect in $G^{\prime\prime}$ a vertex $a_1 \in A\setminus N(B')$ and a vertex $a_2 \in N(B')$ by an edge $e_b \in E(G^{\prime\prime})$ if $a_1, a_2$ share a common neighbor $b$ in the original graph $G$. 
    There is a bijection between the edge-set of $G^{\prime\prime}$ and a subset of the vertex set $B\setminus B'$.
    By Observation \ref{observation:1}, the degree of every vertex $a\in A\setminus N(B')$ in $G^{\prime\prime}$ is equal to $d$.
    \begin{figure}[htbp]
    \centering
    \begin{subfigure}[b]{0.45\textwidth}
    \centering
    \begin{tikzpicture}[
        node distance=1.0cm,
        vertex/.style={circle, draw, minimum size=0.5cm},
        edge/.style={draw}]
        
        \node[vertex, fill=blue!40] (A1) at (0,2) {$a_1$};
        \node[vertex, fill=red!40] (A2) at (0,0.5) {$a_2$};
        \node[vertex, fill=red!40] (A3) at (0,-1) {$a_3$};
        
        \node[vertex] (B1) at (2,3) {$b_1$};
        \node[vertex] (B2) at (2,2) {$b_2$};
        \node[vertex] (B3) at (2,1) {$b_3$};
        \node[vertex] (B4) at (2,0) {$b_4$};
        \node[vertex] (B5) at (2,-1) {$b_5$};
        \node[vertex] (B6) at (2,-2) {$b_6$};
        
        \draw[edge] (B1) -- (A1);
        \draw[edge] (B1) -- (A2);
        \draw[edge] (B2) -- (A1);
        \draw[edge] (B2) -- (A2);
        \draw[edge] (B3) -- (A1);
        \draw[edge] (B3) -- (A3);
        \draw[edge] (B4) -- (A2);
        \draw[edge] (B4) -- (A3);
        \draw[edge] (B5) -- (A1);
        \draw[edge] (B5) -- (A3);
        \draw[edge] (B6) -- (A2);
        \draw[edge] (B6) -- (A3);
    \end{tikzpicture}
    \caption{Graph $G$ with $B^\prime = \{b_6\}$}
    \end{subfigure}
    \hfill
    \begin{subfigure}[b]{0.45\textwidth}
    \centering
    \begin{tikzpicture}[
        node distance=1.0cm,
        vertex/.style={circle, draw, minimum size=0.5cm},
        edge/.style={draw}]
        
        \node[vertex, fill=blue!40] (A1) at (0,0.5) {$a_1$};
        \node[vertex, fill=red!40] (A2) at (2,2.5) {$a_2$};
        \node[vertex, fill=red!40] (A3) at (2,-1.5) {$a_3$};

        \draw[edge] (A1) to[bend left=15] node[left] {$e_{b_1}$} (A2);
        \draw[edge] (A1) to[bend right=15] node[right] {$e_{b_2}$} (A2);
        \draw[edge] (A1) to[bend left=15] node[right] {$e_{b_5}$} (A3);
        \draw[edge] (A1) to[bend right=15] node[left] {$e_{b_3}$} (A3);

    \end{tikzpicture}
    \caption{The auxiliary bipartite graph $G^{\prime\prime}$ on $A\setminus N(B^\prime)=\set{a_1}$ and $N(B^\prime) = \{a_2, a_3\}$}
    \end{subfigure}
    \caption{An example of the construction of the auxiliary graph $G^{\prime\prime}$ of a $(4,2)$-biregular graph $G$ with $B^\prime=\{b_6\}$. The set $A\setminus N(B^\prime)$ is colored in blue and the set $N(B^\prime)$ is colored in red.}
    \label{fig:construction-G-prime-prime}
    \end{figure}
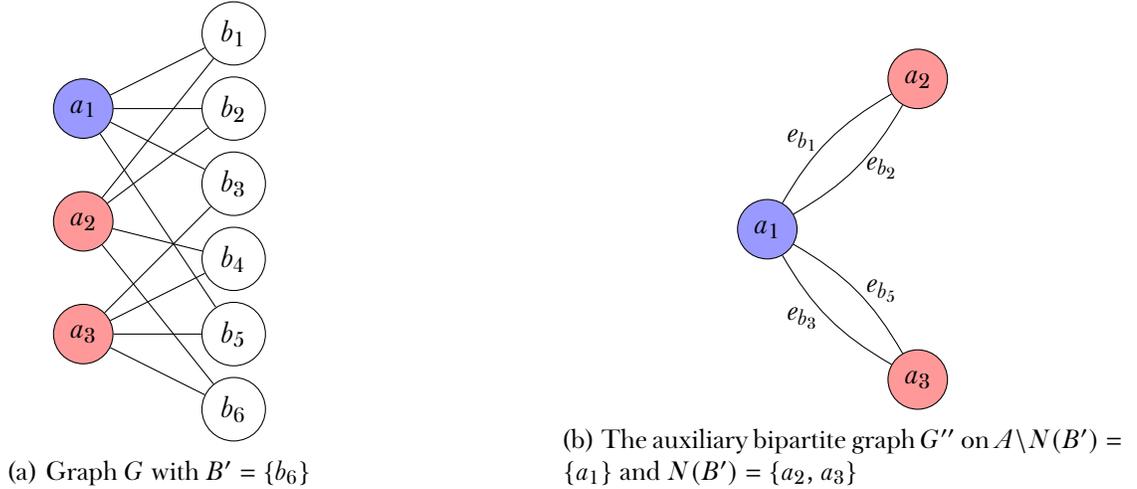
    
    \bigskip
    We claim that there exists an $(A\setminus N(B'))$-perfect matching in $G^{\prime\prime}$. Indeed, this follows from Hall's theorem. Each vertex in $A\setminus N(B')$ has degree $d$ in $G^{\prime\prime}$ while every vertex in $N(B')$ has degree at most $d-1$ in $G^{\prime\prime}$. It is now easy to verify that Hall's condition holds for the graph $G^{\prime\prime}$ and any subset of $A\setminus N(B')$.
    
    Let $M^{\prime\prime}$ be such an $(A\setminus N(B'))$-perfect matching in $G^{\prime\prime}$. $M^{\prime\prime}$ can be found in polynomial-time. The matching $M^{\prime\prime}$ corresponds to subset $S^\prime \subseteq B\setminus B'$ of size $|A\setminus N(B')|$ and $N(S^\prime) \supseteq A\setminus N(B')$.
    To conclude the proof we consider $S := B'\cup S^\prime$. 
    After removing $S$ from $G$, the degree of each vertex in $A$ is at least $d-2\geq 2$ (here we use the fact $d \geq 4$). We can now apply Hall's theorem and find an $A$-perfect matching in $G$ between $A$ and $B\setminus S$. This is because the degree of every vertex in $B$ is equal to $2$ and consequently Hall's condition is satisfied in the subgraph of $G$ induced by $A$ and $B \setminus S$.
    
    Because $d\geq 4$, we have $|B|=m = dn/2\geq 2n$. 
    Therefore,
    \[
    |S| = |B'| + |S^\prime| = |B'| + |A\setminus N(B')| = |B'| + n - 2|B'| = n -|B'|\leq n \leq m-n.
    \]
    Therefore $S$ is a feasible solution to \twoppm. 
    Moreover, $N(S) = N(B') \cup N(S^\prime) = A$ in the original graph $G$, hence $S$ is optimal for \twoppm. 
    For \twopm, the same solution $S$ together with $W=A$ is also optimal since $|N(S)\cap W| = |A| = n \geq \text{OPT}_{2PM}$.
    This concludes the proof.
\end{proof}

\subsection{Algorithms for General 2PPM via Submodular Maximization}\label{sec:submodular}
We show that for inputs that do not fall into the degree constrained frameworks shown above, we may approximate \twoppm by a factor $(1-1/e)$ via submodular maximization. Recall that in our framework, we are given a bipartite graph $G = (A \cup B, E)$ for which we would like to find a set $S \subseteq B$ such that the $A\times (B \setminus S)$ contains a perfect matching and we maximize the number of nodes in $A$ with a neighbor in $S$. This constraint on $B \setminus S$ can be formulated with respect to the transversal matroid $\mathcal{M} = (B, \mathcal{I})$ of the bipartite graph $G$. Here, a subset $I \subseteq B$ is independent if it is a set of endpoints of a matching in $G$. 
\begin{definition}
    Given a bipartite graph  $G=(A \cup B, E)$, the \textit{transversal matroid} $\mathcal{M}$ in $G$ is defined as $\mathcal{M} = (B,\mathcal{I})$ where $\mathcal{I}$ are the subsets of $B$ that are sets of endpoints of a matching in $G$,
    \[
        \mathcal{I} = \{I \subseteq B : \exists \text{ matching }M \text{ of } G \text{ such that } I \subseteq v(M)\}.
    \]
\end{definition}
$S$ is a solution to the \textsc{Second Price Perfect Matching} if and only if $B \setminus S$ contains a basis of $\mathcal{M}$, in other words, $B \setminus S$ is a spanning set. Indeed, since $G$ contains an $A$-perfect matching, each basis of the transversal matroid $\mathcal{M}$ is of size $|A|$ and is covered by an $A$-perfect matching. Similarly, if there is an $A$-perfect matching between $A$ and $B \setminus S$, then the endpoints of this matching form a basis in $\mathcal{M}$. Thus, the \textsc{Second Price Perfect Matching} can be reformulated as the following optimization problem,
\begin{equation}\label{formulation-submod}
\begin{aligned}
    \max_{S \subseteq B} \quad & |N(S)|
    \\
    \text{s.t. } \quad 
    & B \setminus S \text{ contains a basis of } \mathcal{M}.
\end{aligned}
\end{equation}
Here, $N(S) \coloneqq \{a \in A: \exists ab \in E, b \in S\}$ is the neighborhood of $S$. The constraint above is captured naturally by matroid duality.
\begin{definition}\label{def: dual matroid}
    Given a matroid $\mathcal{M} = (B, \mathcal{I})$, $S \subseteq B$ is called a \textit{spanning set} if $S$ contains a basis of $\mathcal{M}$. The \textit{dual matroid} of $\mathcal{M}$ is $\mathcal{M}^* = (B, \mathcal{I}^*)$ where the dual independent sets $\mathcal{I}^*$ are defined by 
    \[
    \mathcal{I}^* = \set{S \subseteq B: B \setminus S \;\text{is a spanning set of}\; \mathcal{M}}.
    \]
\end{definition}
Let $\mathcal{M}$ be the transversal matroid in $G$.
By \cref{def: dual matroid}, we can therefore rewrite the \textsc{Second Price Perfect Matching} as,
\begin{equation}\label{formulation-submod}
    \begin{aligned}
    \max_{S \subseteq B} \quad & |N(S)|
    \\
    \text{s.t. } \quad 
    & S \text{ is independent in } \mathcal{M}^*.
    \end{aligned}
\end{equation}
Note that $f(S) = |N(S)|$ is a coverage function and thus, submodular. Indeed, for all $S,T \subseteq B$,
\[
    |N(S)| + |N(T)| \geq |N(S \cup T)| + |N(S \cap T)|.
\]
Thus, there exists a $(1 - 1/e)$ approximation algorithm for general inputs to the \textsc{Second Price Perfect Matching} via maximizing a submodular function over a single matroid constraint \cite{calinescu2007maximizing}.

\subsection{Other Bounded Degree Cases}\label{sec: deg a is 2}
The case where $\deg(a) = 2$ for all $a \in A$ can be solved in polynomial-time as well for \twoppm. Here, the size of $N(S)$ for a solution $S \subseteq B$ is equal to the number of edges between $S$ and $A$. Indeed, each $a \in A$ has at most one neighbor in $S$ since there has to be an $A$-perfect matching between $A$ and $B \setminus S$. In particular, $|N(S)| = \sum_{b \in S} \deg(b)$. Hence, the optimal solution to \ppm is equal to a maximum-weight independent set in $\mathcal{M}^*$ where the weight of each element $b \in B$ is $\deg(b)$. 
\begin{proposition}
    The \ppm problem with input $G = (A \cup B, E)$, $\deg(a) = 2$ for $a \in A$ can be solved exactly in polynomial-time.
\end{proposition}

Furthermore, if the degrees of vertices in $A$ and $B$ differ significantly, any feasible solution to \pm and \ppm is a good approximation to the optimal solutions.

\begin{proposition}
    \label{prop:degrees}
    For a bipartite graph $G=(A\cup B, E)$ where $\deg(a) \geq d_A, \forall\, a\in A$ and $\deg(b) \leq d_B, \forall\, b\in B$, then any $S \subseteq B$ of size $|S| = |B|-|A|$ satisfies $|N(S)| \geq \left(1-\frac{d_B}{d_A}\right)|A|$.
\end{proposition}

\begin{proof}
    Let $B^\prime = B\setminus S$ and $A^\prime = A\setminus N(S)$. Since the neighbors of $A^\prime$ are all in $B^\prime$, we have 
    \[
    |A^\prime| \cdot d_A \leq \sum_{a\in A^\prime} \deg(a) = E(A^\prime, B^\prime) \leq \sum_{b \in B^\prime} \deg(b) \leq |B^\prime| \cdot d_B.
    \]
    Then we get $|A^\prime| \leq |B^\prime|\cdot\frac{d_B}{d_A} = n \cdot \frac{d_B}{d_A} $, hence $|N(S)| \geq n\left(1-\frac{d_B}{d_A}\right)$.
\end{proof}

\begin{observation}
    When $\deg(a) \geq d_A$ and $\deg(b) \leq d_B$, we get an $\left(1-\frac{d_B}{d_A}\right)$ approximation algorithm for \pm and \ppm. By Hall's Theorem, the graph $G$ always contains a $A$-perfect matching. Thus, there exists a feasible solution $S \subseteq B$ and $W = A$ to \pm of size $|S| = |B| - |A|$. Moreover, this solution is a $\left(1-\frac{d_B}{d_A}\right)$ approximation for \pm and \ppm by Proposition \ref{prop:degrees}. 
    When $d_A \geq e\cdot d_B$, this algorithm outperforms the submodular maximization framework for \ppm. When $d_A \geq 2\cdot d_B$, this algorithm outperforms the prior $1/2$ approximation guarantee for \pm.  
\end{observation}

\section{Lower Bounds of Second Price Perfect Matching}\label{sec:lower bounds}
We now turn to discussing the complexity of \twoppm via algorithmic lower bounds. 
Recall that for \twopm, the best known lower bound is APX-hardness shown in \cite{abkn09}. As a comparison, we will prove that a stronger lower bound holds for \twoppm in Section \ref{sec: inapprox}. In particular, we show that \twoppm is hard to approximate within a factor better than $1-1/e$ under the assumption that P $\neq$ NP. 
In Section \ref{sec: apx-hardness} we will prove that \twoppm is APX-hard even on bounded degree inputs.
\subsection{Inapproximability of \textsc{2PPM}}\label{sec: inapprox}
We first prove that the general \textsc{Second Price Perfect Matching} with no degree constraints is at least as hard as the \textsc{Max $k$-Cover}, which as shown in \cite{fei98} is hard to approximate better than $1-1/e$. 
Therefore the $(1-1/e)$-approximation obtained by the monotone submodular maximization framework in Section \ref{sec:submodular} is in fact the best approximation ratio that can be achieved.
\begin{center}
    \begin{algorithm}[H]
    \caption{\;\textsc{Max $k$-Cover}}
    \begin{algorithmic}
        \Require A set $\mathcal{U}$ of $n$ elements $\set{u_1, \cdots, u_n}$, a family 
        $\mathcal{S}$ of $m$ subsets $\set{S_1, \cdots, S_m}$, and an integer $k\leq m$.
        \Ensure A sub-family $\mathcal{T} \subseteq \mathcal{S}$ of size $k$, which covers the maximum number of elements in $\mathcal{U}$.
    \end{algorithmic}
    \end{algorithm}
\end{center}

\inapx*

\begin{proof}
    We will establish a gap-preserving reduction from the \textsc{Max $k$-Cover} problem to the \textsc{Second Price Perfect Matching}. 
    Given an instance $\mathcal{I}$ of the \textsc{Max $k$-Cover} problem, we formulate $\mathcal{I}$ as a bipartite graph as in Figure \ref{fig:max k-cover}: $\mathcal{U} \cup \mathcal{S}$ is the set of vertices; a vertex $u \in \mathcal{U}$ is connected to a vertex $S \in \mathcal{S}$ if $u \in S$.
    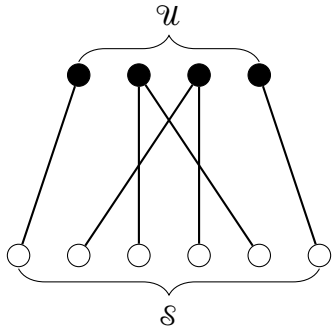
\begin{figure}[ht]
        \centering
        \begin{tikzpicture}[scale=0.8]
        \foreach \i in {1,...,4} {
            \node[circle, draw=black, fill=black, inner sep=3pt] (L\i) at (-\i, 3) {};
        }
        
        \foreach \i in {1,...,6} {
            \node[circle, draw=black, inner sep=3pt] (R\i) at (-\i+1, 0) {};
        }
        
        \draw[black, thick] (L1) -- (R1);
        \draw[black, thick] (L2) -- (R5);
        \draw[black, thick] (L2) -- (R3);
        \draw[black, thick] (L3) -- (R4);
        \draw[black, thick] (L3) -- (R2);
        \draw[black, thick] (L4) -- (R6);
        
        \draw [decorate,decoration={brace,amplitude=10pt,mirror,raise=5pt}] (-1,3) -- (-4,3) node [black,midway,yshift=0.8cm] {$\mathcal{U}$};
    
        \draw [decorate,decoration={brace,amplitude=10pt,raise=5pt}] (-0, 0) -- (-5, 0) node [black,midway,yshift=-0.8cm] {$\mathcal{S}$};
    
        \end{tikzpicture}
        \caption{An instance $\mathcal{I}$ of \textsc{Max $k$-Cover} formulated as a bipartite graph: the part on top represents the set of elements $\mathcal{U}$ and the part at bottom represents the set of subsets $\mathcal{S}$. 
        }
        \label{fig:max k-cover}
    \end{figure}
    
    From the given instance $\mathcal{I}$, we construct an instance $G$ of the \textsc{Second Price Perfect Matching} as follows. 
    First, take $N$ copies of $\mathcal{U}$ where $N \in \NN^+$ is to be determined. Denote those copies 
    as $\mathcal{U}_1, \dots, \mathcal{U}_N$. Each copy $\mathcal{U}_i$ is connected to $\mathcal{S}$ in the same way as $\mathcal{U}$. See Figure \ref{fig:extra-N-copies}.
    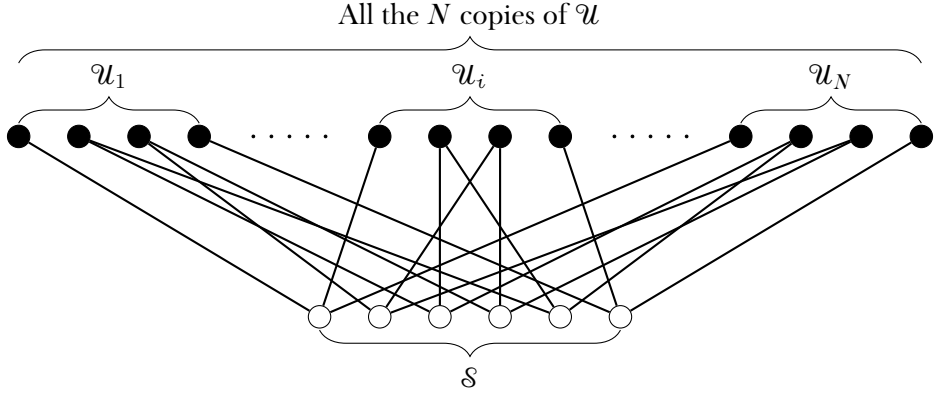
\begin{figure}[ht]
        \centering
        \begin{tikzpicture}[scale=0.8]
        \foreach \i in {1,...,4} {
            \node[circle, draw=black, fill=black, inner sep=3pt] (L\i) at (-\i, 3) {};
        }

        \foreach \i in {1,...,4} {
            \node[circle, draw=black, fill=black, inner sep=3pt] (Lx\i) at (-\i-6, 3) {};
        }

        \foreach \i in {1,...,5} {
            \node[circle, draw=black, fill=black, inner sep=0.3pt] (dx\i) at (-6.4+0.3*\i, 3) {};
        }

        \foreach \i in {1,...,4} {
            \node[circle, draw=black, fill=black, inner sep=3pt] (Ly\i) at (-\i+6, 3) {};
        }

        \foreach \i in {1,...,5} {
            \node[circle, draw=black, fill=black, inner sep=0.3pt] (dx\i) at (-0.4+0.3*\i, 3) {};
        }
        
        \foreach \i in {1,...,6} {
            \node[circle, draw=black, inner sep=3pt] (R\i) at (-\i+1, 0) {};
        }
        
        \draw[black, thick] (L1) -- (R1);
        \draw[black, thick] (L2) -- (R5);
        \draw[black, thick] (L2) -- (R3);
        \draw[black, thick] (L3) -- (R4);
        \draw[black, thick] (L3) -- (R2);
        \draw[black, thick] (L4) -- (R6);

        \draw[black, thick] (Lx1) -- (R1);
        \draw[black, thick] (Lx2) -- (R5);
        \draw[black, thick] (Lx2) -- (R3);
        \draw[black, thick] (Lx3) -- (R4);
        \draw[black, thick] (Lx3) -- (R2);
        \draw[black, thick] (Lx4) -- (R6);
        
        \draw[black, thick] (Ly1) -- (R1);
        \draw[black, thick] (Ly2) -- (R5);
        \draw[black, thick] (Ly2) -- (R3);
        \draw[black, thick] (Ly3) -- (R4);
        \draw[black, thick] (Ly3) -- (R2);
        \draw[black, thick] (Ly4) -- (R6);
         
        \draw [decorate,decoration={brace,amplitude=10pt,mirror,raise=5pt}] (-1,3) -- (-4,3) node [black,midway,yshift=0.8cm] {$\mathcal{U_i}$};

        \draw [decorate,decoration={brace,amplitude=10pt,mirror,raise=5pt}] (-7,3) -- (-10,3) node [black,midway,yshift=0.8cm] {$\mathcal{U_1}$};

        \draw [decorate,decoration={brace,amplitude=10pt,mirror,raise=5pt}] (5,3) -- (2,3) node [black,midway,yshift=0.8cm] {$\mathcal{U}_N$};

        \draw [decorate,decoration={brace,amplitude=10pt,mirror,raise=5pt}] (5,4) -- (-10,4) node [black,midway,yshift=0.8cm] {All the $N$ copies of $\mathcal{U}$};
    
        \draw [decorate,decoration={brace,amplitude=10pt,raise=5pt}] (-0, 0) -- (-5, 0) node [black,midway,yshift=-0.8cm] {$\mathcal{S}$};
    
        \end{tikzpicture}
        \caption{$N$ copies of $\mathcal{U}$ all with the same connections to $\mathcal{S}$.}
        \label{fig:extra-N-copies}
    \end{figure}
    
    Then for each $u$ in all $N$ copies of $\mathcal{U}$, create a ``private'' node 
    which connects only to the corresponding $u$. For each copies $\mathcal{U}_i$, denote its corresponding private nodes as $P_i$.
    Eventually, add $m-k$ dummy nodes, denoted as $D$, 
    and connect each of them to every element of $\mathcal{S}$. 
    The final constructed bipartite graph $G = (A\cup B, E)$ is illustrated in Figure \ref{fig:final-construction}, where $A=\mathcal{U}_1\cup\cdots\cup \mathcal{U}_N \cup D$ and $B=P_1\cup\cdots\cup P_N \cup \mathcal{S}$. The edges $E$ are described above. Note that, 
    \[
    |A| = N\cdot n + (m-k) \leq N\cdot n + m = |B|.
    \]

    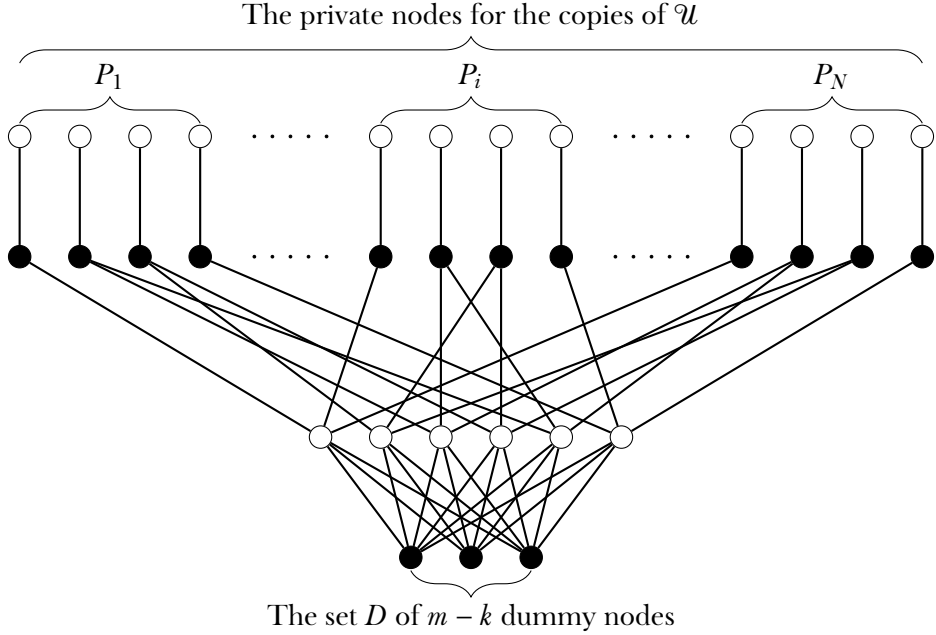
\begin{figure}[ht]
        \centering
        \begin{tikzpicture}[scale=0.8]
        \foreach \i in {1,...,4} {
            \node[circle, draw=black, fill=black, inner sep=3pt] (L\i) at (-\i, 3) {};
        }
        \foreach \i in {1,...,4} {
            \node[circle, draw=black, inner sep=3pt] (LL\i) at (-\i, 5) {};
        }
        \foreach \i in {1,...,4} {
            \node[circle, draw=black, fill=black, inner sep=3pt] (Lx\i) at (-\i-6, 3) {};
        }
        \foreach \i in {1,...,4} {
            \node[circle, draw=black, inner sep=3pt] (LLx\i) at (-\i-6, 5) {};
        }
        \foreach \i in {1,...,5} {
            \node[circle, draw=black, fill=black, inner sep=0.3pt] (dx\i) at (-6.4+0.3*\i, 3) {};
        }
        \foreach \i in {1,...,5} {
            \node[circle, draw=black, fill=black, inner sep=0.3pt] (ddx\i) at (-6.4+0.3*\i, 5) {};
        }
        \foreach \i in {1,...,4} {
            \node[circle, draw=black, fill=black, inner sep=3pt] (Ly\i) at (-\i+6, 3) {};
        }
        \foreach \i in {1,...,4} {
            \node[circle, draw=black, inner sep=3pt] (LLy\i) at (-\i+6, 5) {};
        }
        \foreach \i in {1,...,5} {
            \node[circle, draw=black, fill=black, inner sep=0.3pt] (dy\i) at (-0.4+0.3*\i, 3) {};
        }    
        \foreach \i in {1,...,5} {
            \node[circle, draw=black, fill=black, inner sep=0.3pt] (ddy\i) at (-0.4+0.3*\i, 5) {};
        }
        \foreach \i in {1,...,6} {
            \node[circle, draw=black, inner sep=3pt] (R\i) at (-\i+1, 0) {};
        }
        \foreach \i in {1,...,3} {
            \node[circle, draw=black, fill=black, inner sep=3pt] (D\i) at (\i-4.5, -2) {};
        }
        \draw[black, thick] (L1) -- (R1);
        \draw[black, thick] (L2) -- (R5);
        \draw[black, thick] (L2) -- (R3);
        \draw[black, thick] (L3) -- (R4);
        \draw[black, thick] (L3) -- (R2);
        \draw[black, thick] (L4) -- (R6);

        \draw[black, thick] (Lx1) -- (R1);
        \draw[black, thick] (Lx2) -- (R5);
        \draw[black, thick] (Lx2) -- (R3);
        \draw[black, thick] (Lx3) -- (R4);
        \draw[black, thick] (Lx3) -- (R2);
        \draw[black, thick] (Lx4) -- (R6);
        
        \draw[black, thick] (Ly1) -- (R1);
        \draw[black, thick] (Ly2) -- (R5);
        \draw[black, thick] (Ly2) -- (R3);
        \draw[black, thick] (Ly3) -- (R4);
        \draw[black, thick] (Ly3) -- (R2);
        \draw[black, thick] (Ly4) -- (R6);

        \foreach \i in {1,...,4} {
            \draw[black, thick] (L\i) -- (LL\i);
            \draw[black, thick] (Lx\i) -- (LLx\i);
            \draw[black, thick] (Ly\i) -- (LLy\i);
        }
        \foreach \i in {1,...,6} {
            \foreach \j in {1, ..., 3}{
                \draw[black, thick] (R\i) -- (D\j);
            }
        }

        \draw [decorate,decoration={brace,amplitude=10pt,mirror,raise=5pt}] (-1,5) -- (-4,5) node [black,midway,yshift=0.8cm] {$P_i$};
        \draw [decorate,decoration={brace,amplitude=10pt,mirror,raise=5pt}] (-7,5) -- (-10,5) node [black,midway,yshift=0.8cm] {$P_1$};
        \draw [decorate,decoration={brace,amplitude=10pt,mirror,raise=5pt}] (5,5) -- (2,5) node [black,midway,yshift=0.8cm] {$P_N$};
        \draw [decorate,decoration={brace,amplitude=10pt,mirror,raise=5pt}] (5,6) -- (-10,6) node [black,midway,yshift=0.8cm] {The private nodes for the copies of $\mathcal{U}$};
        \draw [decorate,decoration={brace,amplitude=10pt,raise=5pt}] (-1.5, -2) -- (-3.5, -2) node [black,midway,yshift=-0.8cm] {The set $D$ of $m-k$ dummy nodes};
    
        \end{tikzpicture}
        \caption{Add a ``private" node for each $u$ in all $N$ copies of $\mathcal{U}$ and $m-k$ dummies connected completely to $\mathcal{S}$. 
        }
        \label{fig:final-construction}
    \end{figure}
    
    The following claim shows the relation between the optimal values of the two instances $\mathcal{I}$ and $G$.
    \begin{claim}\label{claim:inapprox}
    Let $\text{OPT}_{MC}$ be the optimal value of $\mathcal{I}$ and let $\text{OPT}_{2PPM}$ be the optimal value of $G$. Then we have
          \begin{equation}
              \text{OPT}_{2PPM} = N\cdot\text{OPT}_{MC} + (m-k).
          \end{equation}  
    \end{claim}

    \begin{cproof}
        We first prove that $\text{OPT}_{2PPM} \geq N\cdot\text{OPT}_{MC} + (m-k)$. 
        Start with an optimal solution of the \textsc{Max $k$-Cover} instance $\mathcal{I}$, which is a sub-collection $\mathcal{T} \subseteq \mathcal{S}$ of size $k$. 
        Construct the $A$-perfect matching by matching each node in $\mathcal{U}_i$ to its corresponding ``private'' node in $P_i$, for all $i=1,\dots, n$, and matching the dummy nodes in $D$ in any fixed way to nodes in $\mathcal{S} \setminus \mathcal{T}$.
        Then in each copy $\mathcal{U}_i$, there are $\text{OPT}_{MC}$ nodes who get covered by $\mathcal{T}$. Also all the $m-k$ dummy nodes in $D$ are covered by $\mathcal{T}$. Hence the total number of nodes in $A$ covered by $\mathcal{T}$ is $N\cdot\text{OPT}_{MC} + (m-k)$. 
        
        We then prove that $\text{OPT}_{2PPM} \leq N\cdot\text{OPT}_{MC} + (m-k)$. 
        Let $M$ be the $A$-perfect matching which is optimal, i.e., the unsaturated nodes in $B$ cover $\text{OPT}_{2PPM}$ many nodes in $A$. 
        First we can assume that in the matching $M$, each node $u\in \mathcal{U}_i$ for any $i$ is matched to its corresponding private node in $P_i$, since otherwise by rematching $u$ to its private node we will not decrease the number of nodes in $A$ which are covered by the unsaturated nodes in $M$. 
        Then for the dummy nodes in $D$, all of them are covered by the unsaturated nodes in $M$. Therefore, the unsaturated $k$ nodes in $\mathcal{S}$ cover $(\text{OPT}_{2PPM} - (m-k))/N$ nodes in each $\mathcal{U}_i$, i.e., it forms a $k$-cover which covers $(\text{OPT}_{2PPM} - (m-k))/N$ many nodes in $\mathcal{U}$, which implies that 
        \[
        \frac{\text{OPT}_{2PPM} - (m-k)}{N} \leq \text{OPT}_{MC}.
        \]
\end{cproof}

Now we finish the proof of Theorem \ref{thm: 1-1/e-inapprox}.
By Claim \ref{claim:inapprox}, for any $\varepsilon>0$ the reduction satisfies
\begin{itemize}
    \item if $\text{OPT}_{MC} \geq s$, then $\text{OPT}_{2PPM} \geq Ns + (m-k)$, 
    \item if $\text{OPT}_{MC} \leq \rb{1-\frac{1}{e}+\varepsilon}s$, then $\text{OPT}_{2PPM} \leq \rb{1-\frac{1}{e}+\varepsilon}Ns + (m-k)$.
\end{itemize}
As shown in \cite{fei98}, the \textsc{Max $k$-Cover} problem cannot be approximated better than $(1-1/e)$ assuming P $\neq$ NP, i.e., it is NP-hard to distinguish between $\text{OPT}_{MC} \geq s$ and $\text{OPT}_{MC} \leq \rb{1-1/e+\varepsilon}s$. Then the reduction implies that it is NP-hard to distinguish between $\text{OPT}_{2PPM} \geq Ns + (m-k)$ and $\text{OPT}_{2PPM} \leq \rb{1-1/e+\varepsilon}Ns + (m-k)$. Note that for any $c > 1-1/e+\varepsilon$, we can find an integer $N$ large enough, such that the ratio
\[
\frac{\rb{1-\frac{1}{e}+\varepsilon}Ns + (m-k)}{Ns + (m-k)} < c,
\]
hence we conclude that it is NP-hard to approximate the \textsc{Second Price Perfect Matching} within a factor $\rb{1-1/e+\varepsilon}$ for any $\varepsilon>0$.
    
\end{proof}
\subsection{APX-Hardness of Degree-Constrained \textsc{2PPM}}\label{sec: apx-hardness}
We will prove in this section that \ppm remains hard to approximate {even with degree constrained input}.
\apxhard*
\begin{proof}
    We reduce from the \textsc{Vertex Cover} problem on 3-regular graphs, which is NP-hard to approximate to within 100/99 as shown in \cite{cc06}. From an instance of 3-regular \textsc{Vertex Cover} $G=(V, E)$, we construct an instance of the degree-constrained \textsc{Second Price Perfect Matching}, which is a bipartite graph $G^\prime = (A\cup B, E^\prime)$ as follows. The construction is inspired by a reduction shown in \cite{abkn09}.    
    First, for each edge $e \in E$, create a node $a_e \in A$ and a unique ``private" node $b_e \in B$ such that $b_e$ connects only to $a_e$. 
    For each vertex $v \in V$, create a node $b_v \in B$. The node $b_v$ is connected to $a_e$ in $G^\prime$ if and only if $v$ is incident to $e$ in $G$.
    Furthermore, we create a gadget for each node $b_v$, which is a path $P_v: b_v - a_v^1 - b_v^1 - a_v^2 - b_v^2$ where $a_v^1, a_v^2 \in A$ and $b_v^1, b_v^2 \in B$.
    Let $|V|=n$ and $|E|=m$. Then in the constructed instance $G^\prime$, we have 
    \[
    A = \rb{\cup_{i=1}^m a_{e_i}} \cup \rb{\cup_{j=1}^n a_{v_j}^1} \cup \rb{\cup_{j=1}^n a_{v_j}^2} \quad \text{and}\quad B = \rb{\cup_{i=1}^m b_{e_i}} \cup \rb{\cup_{j=1}^n b_{v_j}} \cup \rb{\cup_{j=1}^n b_{v_j}^1} \cup \rb{\cup_{j=1}^n b_{v_j}^2},
    \]
    and $|A|=m+2n$ and $|B|=m+3n$. For the degrees of nodes, we have
    \begin{align*}
        \deg(a_{e}) = 3, \forall\,e\in E \quad &\text{and} \quad \deg(a_{v}^1) = \deg(a_{v}^2) = 2, \forall\,v\in V \\
        \deg(b_{e}) = 1, \forall\,e\in E \quad &\text{and} \quad \deg(b_{v})= 4, \deg(b_{v}^1) =2, \deg(b_{v}^2) = 1, \forall\,v\in V. 
    \end{align*}
    To illustrate the construction, we take $G=K_4$ as an example, which is shown in Figure \ref{fig:vc-3-regular}. The corresponding bipartite graph $G^\prime$ is presented in Figure \ref{fig:apx-hardness}.
    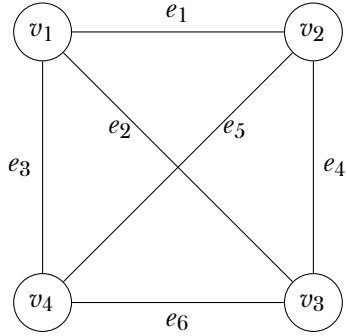
\begin{figure}[htbp]
        \centering
        \begin{tikzpicture}[scale=0.9]
        \begin{scope}[every node/.style={draw, circle, minimum size=1.1em}]
            \node (v1) at (0, 4) {$v_1$};
            \node (v2) at (4, 4) {$v_2$};
            \node (v3) at (4, 0) {$v_3$};
            \node (v4) at (0, 0) {$v_4$};
        \end{scope}
        \draw (v1) -- (v2) node[midway, above] {$e_1$};
        \draw (v1) -- (v3) node[pos=0.25, below] {$e_2$};
        \draw (v1) -- (v4) node[midway, left] {$e_3$};
        \draw (v2) -- (v3) node[midway, right] {$e_4$};
        \draw (v2) -- (v4) node[pos=0.25, below] {$e_5$};
        \draw (v3) -- (v4) node[midway, below] {$e_6$};
        \end{tikzpicture}
        \caption{An instance $G=K_4$ of \textsc{Vertex Cover} problem on 3-regular graphs.}
        \label{fig:vc-3-regular}
    \end{figure}

    \begin{figure}[htbp]
    \centering
    \begin{tikzpicture}[every node/.style={draw, circle, minimum size=1.1em, inner sep=1pt},
    A/.style={fill=blue!40}, B/.style={fill=red!40}, scale=1.2, >=latex]

    
    \node[B] (bv1) at (0, 3.5) {$b_{v_1}$};
    \node[B] (bv2) at (0, 2.2) {$b_{v_2}$};
    \node[B] (bv3) at (0, 0.9) {$b_{v_3}$};
    \node[B] (bv4) at (0, -0.4) {$b_{v_4}$};
    
    \node[A] (av1_1) [right=0.8cm of bv1] {$a_{v_1}^1$};
    \node[B] (bv1_1) [right=0.8cm of av1_1] {$b_{v_1}^1$};
    \node[A] (av1_2) [right=0.8cm of bv1_1] {$a_{v_1}^2$};
    \node[B] (bv1_2) [right=0.8cm of av1_2] {$b_{v_1}^2$};
    
    \draw (bv1) -- (av1_1) -- (bv1_1) -- (av1_2) -- (bv1_2);
    
    \node[A] (av2_1) [right=0.8cm of bv2] {$a_{v_2}^1$};
    \node[B] (bv2_1) [right=0.8cm of av2_1] {$b_{v_2}^1$};
    \node[A] (av2_2) [right=0.8cm of bv2_1] {$a_{v_2}^2$};
    \node[B] (bv2_2) [right=0.8cm of av2_2] {$b_{v_2}^2$};
    
    \draw (bv2) -- (av2_1) -- (bv2_1) -- (av2_2) -- (bv2_2);
    
    \node[A] (av3_1) [right=0.8cm of bv3] {$a_{v_3}^1$};
    \node[B] (bv3_1) [right=0.8cm of av3_1] {$b_{v_3}^1$};
    \node[A] (av3_2) [right=0.8cm of bv3_1] {$a_{v_3}^2$};
    \node[B] (bv3_2) [right=0.8cm of av3_2] {$b_{v_3}^2$};
    
    \draw (bv3) -- (av3_1) -- (bv3_1) -- (av3_2) -- (bv3_2);
    
    \node[A] (av4_1) [right=0.8cm of bv4] {$a_{v_4}^1$};
    \node[B] (bv4_1) [right=0.8cm of av4_1] {$b_{v_4}^1$};
    \node[A] (av4_2) [right=0.8cm of bv4_1] {$a_{v_4}^2$};
    \node[B] (bv4_2) [right=0.8cm of av4_2] {$b_{v_4}^2$};
    
    \draw (bv4) -- (av4_1) -- (bv4_1) -- (av4_2) -- (bv4_2);
    
    \foreach \i in {1,...,6} {
      \node[A] (ae\i) at (-3, 5 - \i) {$a_{e_\i}$};
      \node[B] (be\i) at (-4.5, 5 - \i) {$b_{e_\i}$};
      \draw (be\i) -- (ae\i);
    }
    
    
    \draw (bv1) -- (ae1);
    \draw (bv1) -- (ae2);
    \draw (bv1) -- (ae3);
    
    \draw (bv2) -- (ae1);
    \draw (bv2) -- (ae4);
    \draw (bv2) -- (ae5);
    
    \draw (bv3) -- (ae2);
    \draw (bv3) -- (ae4);
    \draw (bv3) -- (ae6);
    
    \draw (bv4) -- (ae3);
    \draw (bv4) -- (ae5);
    \draw (bv4) -- (ae6);
    \end{tikzpicture}
    
    \caption{The construction of instance $G^\prime$ of \textsc{Second Price Perfect Matching}, based on the instance $G=K_4$ of 3-regular \textsc{Vertex Cover} shown in Figure \ref{fig:vc-3-regular}.}
    \label{fig:apx-hardness}
\end{figure}
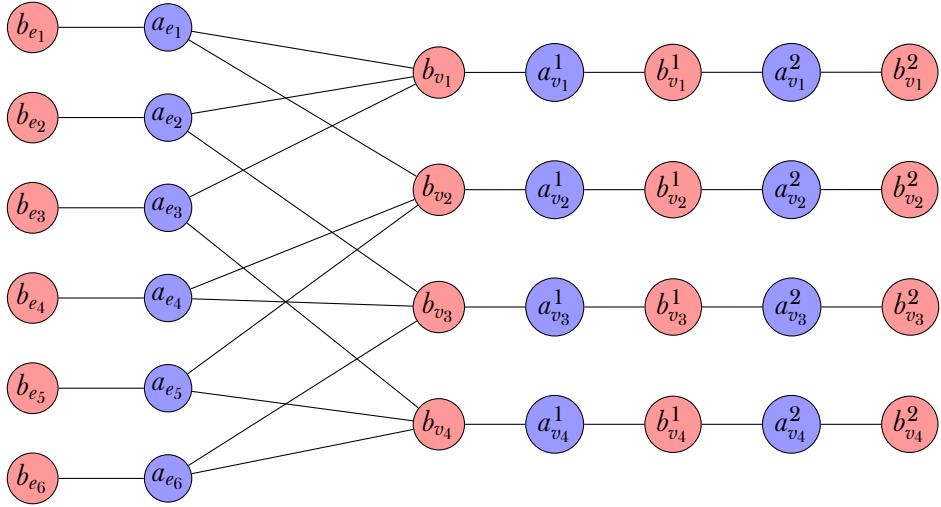
The following lemma shows the relation between optimal values of these two instances.
\begin{claim}\label{claim:apx-hardness}
Let $\text{OPT}_{VC}$ be the size of the minimum vertex cover of $G$ and let $\text{OPT}_{2PPM}$ be the optimal value of \textsc{2PPM} with input $G^\prime$. 
Then we have
\begin{equation}
    \text{OPT}_{2PPM} + \text{OPT}_{VC} = 2n + m.
\end{equation}
\end{claim}
\begin{cproof}
    We start by proving $\text{OPT}_{2PPM} + \text{OPT}_{VC} \geq 2n + m$. 
    Let $U\subseteq V$ be an optimal vertex cover of $G$. We construct a solution to the \textsc{Second Price Perfect Matching} as follows. First we match every $a_{e}$ to its private node $b_{e}$. For the nodes $a_{v}^1, a_{v}^2$ in each path gadget $b_v - a_v^1 - b_v^1 - a_v^2 - b_v^2$ corresponding to $v$, if $v$ is in the vertex cover $U$, then we match $a_{v}^1$ to $b_{v}^1$ and match $a_{v}^2$ to $b_{v}^2$; otherwise, we match $a_{v}^1$ to $b_{v}$ and match $a_{v}^2$ to $b_{v}^2$. Thus we get a $A$-perfect matching. 
    The number of nodes in $A$ which are covered by the unsaturated nodes in $B$ is then $m+1\cdot\text{OPT}_{VC}+2\cdot(n-\text{OPT}_{VC})$, which equals to $2n+m-\text{OPT}_{VC}$. 
    
    Next we prove that $\text{OPT}_{2PPM} + \text{OPT}_{VC} \leq 2n + m$. 
    Let $S \subseteq B$ be any feasible solution to the \textsc{Second Price Perfect Matching} and let $M$ be a $A$-perfect matching between $A$ and $B\setminus S$. 
    We will introduce a transformation from $S$ and $M$ to $S^\prime \subseteq B$ and a $A$-perfect matching $M^\prime$ between $A$ and $B\setminus S^\prime$, such that after the transformation, there will be no less number of nodes in $A$ that are covered by the unsaturated nodes in $B$, i.e., $f(S^\prime) \geq f(S)$; moreover, $\set{v\in V: b_v \;\text{is unsaturated by}\;M^\prime}$ will form a vertex cover of $G$. 
    
    The transformation from $S, M$ to $S^\prime, M^\prime$ works as follows.
    First we match each node $a_{e}$ to its private node $b_{e}$ in $M^\prime$ if $a_e, b_e$ are not matched in $M$. By doing this, there will be no less number of $A$ nodes that are covered by the unsaturated nodes in $M^\prime$ than in $M$. 
    Next for the path gadget $P_v$ of each $v\in V$, we match $a_v^2$ to $b_v^2$ in $M^\prime$ if $a_v^2, b_v^2$ are not matched in $M$. Again by doing this, the number of $A$ nodes which are covered by the unsaturated nodes in $M^\prime$ can only increase or remain the same, compared to $M$. 
    
    Finally in the last step of the transformation we do the following: For each edge $e=uv \in E$ such that both $u$ and $v$ are saturated by $M$, i.e., in the matching $M$, $b_u$ is matched to $a^1_u$ and $b_v$ is matched to $a^1_v$, we make one of $b_u, b_v$ unsaturated in $M^\prime$. For example we may choose $b_u$, and match $a_u^1$ to $b_u^1$ in $M^\prime$, as illustrated in Figure \ref{fig:apx-hardness-transformation}.
    Notice that this will not decrease the number of $A$ nodes that are covered by the unsaturated nodes. Indeed, let us consider the $A$ nodes $a_e, a_u^1, a_u^2, a_v^1, a_v^2$. Before in $M$, there are 4 nodes $a_u^1, a_u^2, a_v^1, a_v^2$ that are covered by the unsaturated nodes whereas $a_e$ is not covered by the unsaturated nodes. Now in $M^\prime$, there are again 4 nodes $a_e, a_u^1, a_v^1, a_v^2$ that are covered by the unsaturated nodes whereas $a_u^2$ is not covered by the unsaturated nodes. 
    Furthermore, there are two other nodes $a_f, a_g$ in $A$ that are adjacent to $b_u$ if $u$ is incident with $f,g \in E$. Since $b_u$ is saturated in $M$ and unsaturated in $M^\prime$, $a_f, a_g$ are covered by the unsaturated nodes in $M^\prime$ and may not be covered by the unsaturated nodes in $M$.
    \begin{figure}[htbp]
        \centering
        \begin{minipage}[b]{1.0\textwidth}
            \centering
            \begin{tikzpicture}[every node/.style={draw, circle, minimum size=1.1em, inner sep=1pt},
            A/.style={fill=blue!20}, B/.style={fill=red!20}, scale=1.2, >=latex]
        
            \node[B] (be) at (-3, 0.5) {$b_{e}$};
            \node[A] (ae) at (-1.5, 0.5) {$a_{e}$};
            
            \node[B] (bu) at (0, 1) {$b_{u}$};
            \node[B] (bv) at (0, 0) {$b_{v}$};
    
            \draw[color=red, line width=3.0pt] (be) -- (ae);
            \draw (ae) -- (bu);
            \draw (ae) -- (bv);
    
            \node[A] (au_1) [right=0.8cm of bu] {$a_{u}^1$};
            \node[B] (bu_1) [right=0.8cm of au_1] {$b_{u}^1$};
            \node[A] (au_2) [right=0.8cm of bu_1] {$a_{u}^2$};
            \node[B] (bu_2) [right=0.8cm of au_2] {$b_{u}^2$};

            \draw[color=red, line width=3.0pt] (bu) -- (au_1);
            \draw (au_1) -- (bu_1) -- (au_2);
            \draw[color=red, line width=3.0pt] (au_2) -- (bu_2);
                    
            \node[A] (av_1) [right=0.8cm of bv] {$a_{v}^1$};
            \node[B] (bv_1) [right=0.8cm of av_1] {$b_{v}^1$};
            \node[A] (av_2) [right=0.8cm of bv_1] {$a_{v}^2$};
            \node[B] (bv_2) [right=0.8cm of av_2] {$b_{v}^2$};

            \draw[color=red, line width=3.0pt] (bv) -- (av_1);
            \draw (av_1) -- (bv_1) -- (av_2);
            \draw[color=red, line width=3.0pt] (av_2) -- (bv_2);
    
            \end{tikzpicture}
            \caption*{(a). Before transformation, the edges in $M$ are colored in red.}
        \end{minipage}%
        \vfill
        \begin{minipage}[b]{1.0\textwidth}
            \centering
            \begin{tikzpicture}[every node/.style={draw, circle, minimum size=1.1em, inner sep=1pt},
            A/.style={fill=blue!20}, B/.style={fill=red!20}, scale=1.2, >=latex]
        
            \node[B] (be) at (-3, 0.5) {$b_{e}$};
            \node[A] (ae) at (-1.5, 0.5) {$a_{e}$};
            
            \node[B] (bu) at (0, 1) {$b_{u}$};
            \node[B] (bv) at (0, 0) {$b_{v}$};
    
            \draw[color=blue, line width=3.0pt] (be) -- (ae);
            \draw (ae) -- (bu);
            \draw (ae) -- (bv);
    
            \node[A] (au_1) [right=0.8cm of bu] {$a_{u}^1$};
            \node[B] (bu_1) [right=0.8cm of au_1] {$b_{u}^1$};
            \node[A] (au_2) [right=0.8cm of bu_1] {$a_{u}^2$};
            \node[B] (bu_2) [right=0.8cm of au_2] {$b_{u}^2$};
            
            \draw (bu) -- (au_1);
            \draw[color=blue, line width=3.0pt] (au_1) -- (bu_1);
            \draw (bu_1) -- (au_2);
            \draw[color=blue, line width=3.0pt] (au_2) -- (bu_2);
            
            \node[A] (av_1) [right=0.8cm of bv] {$a_{v}^1$};
            \node[B] (bv_1) [right=0.8cm of av_1] {$b_{v}^1$};
            \node[A] (av_2) [right=0.8cm of bv_1] {$a_{v}^2$};
            \node[B] (bv_2) [right=0.8cm of av_2] {$b_{v}^2$};
            
            \draw[color=blue, line width=3.0pt] (bv) -- (av_1);
            \draw (av_1) -- (bv_1) -- (av_2);
            \draw[color=blue, line width=3.0pt] (av_2) -- (bv_2);
    
            \end{tikzpicture}
            \caption*{(b). After transformation, the edges in $M^\prime$ are colored in blue.}
        \end{minipage}%
        \caption{The last step of transformation from $M$ to $M^\prime$.}
        \label{fig:apx-hardness-transformation}
    \end{figure}
    
    By applying the transformation described above to an optimal solution $S\subseteq B$ of \textsc{Second Price Perfect Matching}, we get a vertex cover $U$ of $G$. The nodes in $A$ that are covered by the unsaturated nodes in $B$ are 
    \[
    \set{a_e: e\in E} \cup \set{a^1_u: u \in U} \cup \set{a^1_v, a^2_v: v\not\in U},
    \]
    so we have $m + 1\cdot |U| + 2\cdot(n-|U|) = \text{OPT}_{2PPM}$, i.e., $|U|=2n+m-\text{OPT}_{2PPM}$.
\end{cproof}
Now going back to the proof of Theorem \ref{thm: APX-hardness}. Suppose that there is a $\alpha$-approximation algorithm of the \textsc{Second Price Perfect Matching} in polynomial-time, $\alpha \in (0,1]$. 
Let $S \subseteq B$ be the output of the $\alpha$-approximation algorithm running on $G^\prime$. Let $M$ be a $A$-perfect matching between $A$ and $B\setminus S$ and let $f(S)$ be the number of nodes in $A$ which are covered by the unsaturated nodes in $M$. 
First, note that by applying the same transformation as in the proof of Claim \ref{claim:apx-hardness}, we can get $S^\prime \subseteq B$ with $f(S^\prime)\geq f(S)$, such that in the corresponding $A$-perfect matching $M^\prime$, each $a_{e}$ is matched to its ``private" node $b_{e}$ and the unsaturated nodes $b_{v}$ by $M^\prime$ forms a vertex cover of $G$, denoted as $U$. 
We have 
\[
f(S^\prime) \geq f(S) \geq \alpha\cdot \text{OPT}_{2PPM}.
\]
From here we can output a vertex cover $U$ of $G$, which is of size $2n+m-f(S^\prime)$.
Since $G$ is 3-regular, we have $3n=2m$ and $3\text{OPT}_{VC} \geq m$, i.e., $n/\text{OPT}_{VC} \leq 1/2$. By applying Claim \ref{claim:apx-hardness}, we know that the size of $U$ satisfies
\[
|U| = 2n+m-f(S^\prime) \leq 2n+m-\alpha\cdot \text{OPT}_{2PPM} = 2n+m-\alpha\rb{2n+m - \text{OPT}_{VC}} = \frac{7}{2}n - \alpha\rb{\frac{7}{2}n - \text{OPT}_{VC}}.
\]
Eventually, the vertex cover $U$ of $G$ gives an approximation ratio $|U|/\text{OPT}_{VC}$ at least $100/99$ by the hardness result shown in \cite{cc06}, which implies the following
\begin{equation}
    \frac{100}{99} \leq \frac{|U|}{\text{OPT}_{VC}} \leq \frac{\frac{7}{2}n - \alpha\rb{\frac{7}{2}n - \text{OPT}_{VC}}}{\text{OPT}_{VC}} = \frac{7}{2}(1-\alpha)\frac{n}{\text{OPT}_{VC}} + \alpha \leq \frac{7}{4}(1-\alpha) + \alpha,
\end{equation}
hence by direct computations we get $\alpha \leq 293/297$.
\end{proof}


\section*{Acknowledgement}
The authors are grateful to Friedrich Eisenbrand, Anupam Gupta, Thomas Rothvoss, and Theophile Thiery for the helpful discussions and comments.

\newpage
\bibliography{refs}

\begin{thebibliography}{10}

\bibitem{aggarwal2011online}
Gagan Aggarwal, Gagan Goel, Chinmay Karande, and Aranyak Mehta.
\newblock Online vertex-weighted bipartite matching and single-bid budgeted allocations.
\newblock In {\em Proceedings of the twenty-second annual ACM-SIAM symposium on Discrete Algorithms}, pages 1253--1264. SIAM, 2011.

\bibitem{andelman2004auctions}
Nir Andelman and Yishay Mansour.
\newblock Auctions with budget constraints.
\newblock In {\em Scandinavian Workshop on Algorithm Theory}, pages 26--38. Springer, 2004.

\bibitem{azar2008improved}
Yossi Azar, Benjamin Birnbaum, Anna~R Karlin, Claire Mathieu, and C~Thach Nguyen.
\newblock Improved approximation algorithms for budgeted allocations.
\newblock In {\em International Colloquium on Automata, Languages, and Programming}, pages 186--197. Springer, 2008.

\bibitem{abkn09}
Yossi Azar, Benjamin Birnbaum, Anna~R. Karlin, and C.~Thach Nguyen.
\newblock On revenue maximization in second-price ad auctions.
\newblock In Amos Fiat and Peter Sanders, editors, {\em Algorithms - ESA 2009}, pages 155--166, Berlin, Heidelberg, 2009. Springer Berlin Heidelberg.

\bibitem{buchbinder2007online}
Niv Buchbinder, Kamal Jain, and Joseph Naor.
\newblock Online primal-dual algorithms for maximizing ad-auctions revenue.
\newblock In {\em European Symposium on Algorithms}, pages 253--264. Springer, 2007.

\bibitem{calinescu2007maximizing}
Gruia Calinescu, Chandra Chekuri, Martin P{\'a}l, and Jan Vondr{\'a}k.
\newblock Maximizing a submodular set function subject to a matroid constraint.
\newblock In {\em International Conference on Integer Programming and Combinatorial Optimization}, pages 182--196. Springer, 2007.

\bibitem{chakrabarty2010approximability}
Deeparnab Chakrabarty and Gagan Goel.
\newblock On the approximability of budgeted allocations and improved lower bounds for submodular welfare maximization and gap.
\newblock {\em SIAM Journal on Computing}, 39(6):2189--2211, 2010.

\bibitem{cc06}
Miroslav Chlebík and Janka Chlebíková.
\newblock Complexity of approximating bounded variants of optimization problems.
\newblock {\em Theoretical Computer Science}, 354(3):320--338, 2006.
\newblock Foundations of Computation Theory (FCT 2003).

\bibitem{edelman2007internet}
Benjamin Edelman, Michael Ostrovsky, and Michael Schwarz.
\newblock Internet advertising and the generalized second-price auction: Selling billions of dollars worth of keywords.
\newblock {\em American economic review}, 97(1):242--259, 2007.

\bibitem{fei98}
Uriel Feige.
\newblock A threshold of $\ln n$ for approximating set cover.
\newblock {\em Journal of the ACM}, 45(4):634–652, July 1998.

\bibitem{fernandes2014second}
Cristina~G Fernandes and Rafael~CS Schouery.
\newblock Second-price ad auctions with binary bids and markets with good competition.
\newblock {\em Theoretical Computer Science}, 540:103--114, 2014.

\bibitem{garg2001approximation}
Rahul Garg, Vijay Kumar, and Vinayaka Pandit.
\newblock Approximation algorithms for budget-constrained auctions.
\newblock In {\em International Workshop on Randomization and Approximation Techniques in Computer Science}, pages 102--113. Springer, 2001.

\bibitem{kalaitzis2016improved}
Christos Kalaitzis.
\newblock An improved approximation guarantee for the maximum budgeted allocation problem.
\newblock In {\em Proceedings of the Twenty-Seventh Annual ACM-SIAM Symposium on Discrete Algorithms}, pages 1048--1066. SIAM, 2016.

\bibitem{khot2005inapproximability}
Subhash Khot, Richard~J Lipton, Evangelos Markakis, and Aranyak Mehta.
\newblock Inapproximability results for combinatorial auctions with submodular utility functions.
\newblock In {\em International Workshop on Internet and Network Economics}, pages 92--101. Springer, 2005.

\bibitem{lehmann2001combinatorial}
Benny Lehmann, Daniel Lehmann, and Noam Nisan.
\newblock Combinatorial auctions with decreasing marginal utilities.
\newblock In {\em Proceedings of the 3rd ACM conference on Electronic Commerce}, pages 18--28, 2001.

\bibitem{StackExchange}
Mattias and Misha Lavrov.
\newblock 3-regular graph maximum matching.
\newblock \url{https://math.stackexchange.com/questions/3410007/3-regular-graph-maximum-matching}.
\newblock Mathematics Stack Exchange, version: 2025-03-27.

\bibitem{mehta2013online}
Aranyak Mehta et~al.
\newblock Online matching and ad allocation.
\newblock {\em Foundations and Trends{\textregistered} in Theoretical Computer Science}, 8(4):265--368, 2013.

\bibitem{mehta2007adwords}
Aranyak Mehta, Amin Saberi, Umesh Vazirani, and Vijay Vazirani.
\newblock Adwords and generalized online matching.
\newblock {\em Journal of the ACM (JACM)}, 54(5):22--es, 2007.

\bibitem{sch03}
Alexander Schrijver.
\newblock {\em Combinatorial Optimization: Polyhedra and Efficiency}.
\newblock Springer Berlin, Heidelberg, 2003.

\bibitem{srinivasan2008budgeted}
Aravind Srinivasan.
\newblock Budgeted allocations in the full-information setting.
\newblock In {\em International Workshop on Approximation Algorithms for Combinatorial Optimization}, pages 247--253. Springer, 2008.

\bibitem{varian2007position}
Hal~R Varian.
\newblock Position auctions.
\newblock {\em international Journal of industrial Organization}, 25(6):1163--1178, 2007.

\bibitem{vondrak2008optimal}
Jan Vondr{\'a}k.
\newblock Optimal approximation for the submodular welfare problem in the value oracle model.
\newblock In {\em Proceedings of the fortieth annual ACM symposium on Theory of computing}, pages 67--74, 2008.

\end{thebibliography}
\bibliographystyle{plain}

\end{document}